\documentclass{aa}
\usepackage{times}
\usepackage[T1]{fontenc}
\usepackage{graphics}

\providecommand{\LyX}{L\kern-.1667em\lower.25em\hbox{Y}\kern-.125emX\@}
\newcommand{\noun}[1]{\textsc{#1}}
\def\footnote{\ifx\protect\@typeset@protect
    \expandafter\SF@@footnote
  \else
    \expandafter\SF@gobble@opt
  \fi
}
\expandafter\def\csname SF@gobble@opt \endcsname{\@ifnextchar[
  \SF@gobble@twobracket
  \@gobble
}
\edef\SF@gobble@opt{\noexpand\protect
  \expandafter\noexpand\csname SF@gobble@opt \endcsname}
\def\SF@gobble@twobracket[#1]#2{}

\newcommand{\mv}{\mbox{$M_{V}$}}

\newcommand{\bv}{\mbox{$B\!-\!V$}}
\newcommand{\vi}{\mbox{$V\!-\!I$}}

\newcommand{\feh}{\mbox{[Fe/H]}}

\newcommand{\Msun}{\mbox{$M_{\odot}$}}

\newcommand{\sub}[1]{\mbox{$_{\rm #1}$}}
\newcommand{\Teff}{\mbox{$T\sub{eff}$}}
\newcommand{\teff}{\mbox{$T\sub{eff}$}}

\newcommand{\changed}{}

\newcommand{\be}{\begin{equation}} 
\newcommand{\ee}{\end{equation}} 
\newcommand{\nw}{\newcommand} 
\nw{\beqn}{\begin{eqnarray}} 
\nw{\eqn}{\end{eqnarray}} 
\nw{\lft}{\Longleftrightarrow} 
\nw{\lb}{\lambda} 
\nw{\f}{\frac} 
\nw{\p}{\partial} 

\begin{document}

\title{Does the mixing length parameter depend on metallicity?}
\subtitle{Further tests of evolutionary sequences using homogeneous databases }

\author{Rossella Palmieri\inst{1} \and 
	Giampaolo Piotto\inst{1} \and 
	Ivo Saviane\inst{2} \and
	L\'eo Girardi\inst{3,1} \and
	Vittorio Castellani\inst{4}
	}

   \offprints{isaviane@eso.org}

\institute{
Dipartimento di Astronomia, Universit\`a di Padova, Vicolo
dell'Osservatorio 2, I-35122 Padova, Italy \\
\email{palmieri,piotto, lgirardi@pd.astro.it} 
        \and 
European Southern Observatory,
3107 Alonso de Cordova, Casilla 19001, Santiago 19, Chile \\
\email{isaviane@eso.org}
	\and
Osservatorio Astronomico di Trieste,  Via G. B. Tiepolo 11,
I-34131 Trieste, Italy \\
\email{lgirardi@ts.astro.it}
	\and
Dipartimento di Fisica, Università di Pisa, Piazza Torricelli 2, 
I-56100 Pisa, Italy \\ 
\email{vittorio@astr18pi.difi.unipi.it}
	}

   \date{Accepted May 15, 2002}

   \abstract{
This paper is a further step in the investigation of the morphology of
the color-magnitude diagram of Galactic globular clusters, and the
fine-tuning of theoretical models, made possible by the recent
observational efforts to build homogeneous photometric databases.  In
particular, we examine here the calibration of the morphological
parameter $W_{\rm HB}$ vs. metallicity, originally proposed by Brocato
et al. (\cite{brocatoEtal98}; B98), which essentially measures the color
position of the red-giant branch. We show that the parameter can be used
to have a first-order estimate of the cluster metallicity, since the
dispersion around the mean trend with [Fe/H] is compatible with the
measurement errors.
The tight $W_{\rm HB}$-[Fe/H] relation is then used to show that
variations in helium content or age do not affect the parameter, whereas
it is strongly influenced by the mixing-length parameter $\alpha$ (as
expected). This fact allows us, for the first time, to state that there
is no trend of $\alpha$ with the metal content of a cluster.
A thorough examination of the interrelated questions of the
$\alpha$-elements enhancement and the color-$T_{\rm eff}$
transformations, highlights that there is an urgent need for an
independent assessment of which of the two presently accepted
metallicity scales is the true indicator of a cluster's iron content.
Whatever scenario is adopted, it also appears that a deep revision 
of the $V-I$-temperature relations is needed.
   }

   \maketitle

\section{Introduction}

\begin{figure}
{\par\centering \resizebox*{0.9\columnwidth}{!}{\includegraphics{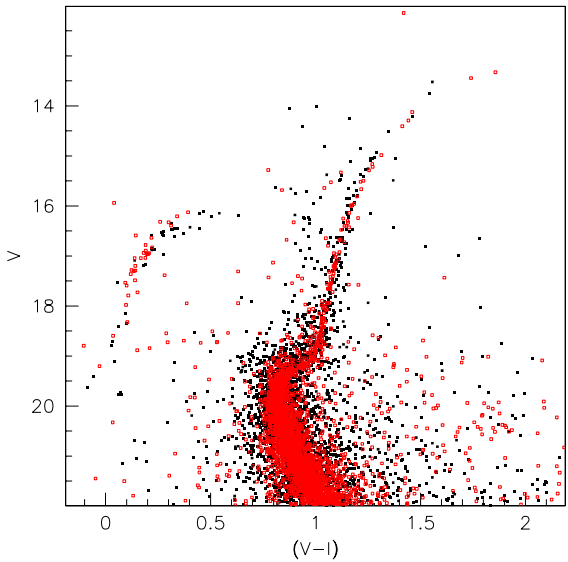}} \par}
\caption{M80 (filled squares) vs. M12 (open squares). An offset $\delta (V-I)=-0.03$
and $\delta V=+1.50$ was applied to the M12 diagram before
overlapping it to that of M80 \label{fig:m80vsm12}}
\end{figure}

The comparison of the observed color-magnitude diagrams (CMD) of a
star cluster with the theoretical isochrones is the best tool we have
to tune some fundamental parameters of the stellar evolutionary
models. Only when we are sure that the input parameters and the input
physics are correct, we can use the model to infer some properties of
the stellar population (like age, helium content, etc.), not otherwise
empirically measurable. Clearly, this is a complex job: on one side we
want to use the cluster stars to tune the models, and on the other we
need to use the models to infer properties of the cluster stellar
population. Not surprisingly, any stellar models must adopt a number of
assumptions, often not directly supported by observational evidence.
One of the most uncertain parameters, strongly affecting the theoretical
location of some branches of the CMD, like the red giant branch, is the
mixing length parameter $\alpha$.  This parameter, in the framework of
the mixing-length theory (MLT; B\"ohm-Vitense \cite{bohm-vitense58}),
determines the efficiency of energy transport by convection in the
outermost layers of a star. For a given stellar luminosity, it also
determines the exact radius of the star, and hence its effective
temperature and colors.
   
It is well known that, in order to reproduce the typical temperatures
and colors of red giant stars, $\alpha$ is required to have a value
between 1.5 and 2.0. Also, a value close to 1.7 is required for
reproducing the solar radius in non-diffusive solar models, whereas
about 1.9 is favored when diffusion is taken into account. The fact that
$\alpha$ is similar for red giants and the Sun 
{\changed
(VandenBerg et al. \cite{vandenberg00}; Alonso et al. \cite{alonsoEtal00})}
 leads to the usual approach of calibrating $\alpha$ by means of a
solar model -- i.e. a model with 1~\Msun\ and solar composition, that is
required to have the solar luminosity and radius at an age of $\sim
4.5$~Gyr.  The same $\alpha$ value is then used to model all stars,
including red giants.  However, there is no theoretical justification
that the same value of $\alpha$ should apply
for any star.

In this paper, we have attempted to investigate which is the best value
of $\alpha$ for low mass (globular cluster) stars and, overall, whether
there is any dependence of $\alpha$ with the cluster metallicity. Our
results cannot be considered definitive; however this paper shows a
possible approach to the problem, and enlightens all the uncertainties
associated to the calibration of the mixing length parameter. Also, this
work is limited to the context of the MLT, which is still a fundamental
approach adopted in most computations of stellar models. However, one
should keep in mind that the MLT is admittedly a very approximate
theory, and that alternative approaches (e.g.\ Canuto \& Mazzitelli
\cite{canutomazzitelli91}; 
Spruit \cite{spruit97}; Ludwig et al. \cite{ludwigEtal99}) have been
suggested.

This project has been stimulated by an investigation carried out by some
of us a few years ago.  In the course of a photometric study of the
Galactic globular cluster (GGC) M80 (Brocato et
al. \cite{brocatoEtal98}, hereafter B98), we compared the
morphological characteristics of its color-magnitude diagram (CMD)
with those of an ensemble of other globulars for which a CMD was
available in the literature. That investigation left an open question
concerning the relative position of the RGB with respect to the HB as
a function of the metallicity.  More specifically, when the CMD of M80
was compared with that of GGCs with similar metallicity, we found a
group of clusters (M3, M13, NGC~7492) whose members had CMDs
overlapping that of M80, and another (M12, NGC~1904, NGC~5897) for
which significant discrepancies were seen. The discrepancies were in
the sense that while the HBs could be overlapped in a satisfactory way,
the RGB fiducial lines showed a dispersion in color, M12 having the
reddest branch (cf Fig. 6 and 7 in B98), as expected if M12
would be more metal rich than indicated in the literature. We also excluded
that the discrepancy could be due to an age difference.

In order to have a more quantitative comparison among a larger sample
of clusters, with different metal content, we devised a new
morphological parameter with the aim of quantitatively measuring the
distance in color of the RGB with respect to some fixed point on the
HB.  We selected as a reference point the so called {}``HB
turn-down{}''(HB-TD), since it has been demonstrated that the location
of this point in the CMD is largely unaffected by the cluster age,
metallicity, and primordial He content (cf. the detailed discussion in
B98).  We fixed the HB-TD at $(B-V)_{0}=0$ and measured the RGB color
at $0.5$ magnitudes brighter than the HB-TD level.  As the position of
the HB-
TD is fixed, $W_{\rm HB}$ just shows the displacement of
the RGB as a function of the metal content (cf. Fig.~10 of B98). We
found that the trend with the metallicity is in the direction expected
from the theoretical models, though with a large dispersion,
particularly evident at intermediate metallicities
($-1.85<${[}Fe/H{]}$<-1.50$), larger than expected on the
basis of the measurement errors.

We were not able to decide whether such an evidence might simply be
the consequence of the errors in the photometric calibration, or a
peculiar distribution in the global metallicity of the clusters, and
deferred further discussion until direct measurements of $\alpha$
elements and/or a database of CMDs with homogeneous calibration would
have been available.

Such database is now available, thanks to the efforts of a number of
the original investigators of B98, and we are now able to re-examine
the question, taking also advantage of the new theoretical
calculations that we have specifically performed for this project.

The paper is organized as follows. The datasets are presented in
Sect.~\ref{sec:datasets}. The measurement procedures and the
corrections for differential reddening are described in
Sect.~\ref{sec:measurements} and \ref{sec:diffredd-correction},
respectively. Sect.~\ref{sec:thewhb-parameter} re-examines the
original question of the trend and dispersion of $W_{\rm HB}$ as a
function of metallicity.

Sect.~\ref{sec:compare-models} deals with the comparison of the
VandenBerg et al. (\cite{vandenberg00}; V00) and the Girardi et
al. (\cite{girardiEtal00}; G00) theoretical isochrones to the data.
Due to the lower degree of sampling of the G00 isochrones, a more
detailed description of the turn-down identification and its measurement
is offered in Sect.~\ref{sec:isoc-g00}.

The keys to the interpretation of the observed vs. computed trend of
$W_{\rm HB}$ are offered in Sect.~\ref{sec:what-det-whb}. In
particular, the dependence on metallicity, mixing length, age, helium
content, $\alpha$-enhancement, and the \Teff-color transformations are
examined in Sects.~\ref{sec:what-met},
\ref{sec:what-mixlen},
\ref{sec:what-age},
\ref{sec:what-helium},
\ref{sec:what-enha}, and
\ref{sec:what-transf}.

Whether our data can be used  to constrain
the mixing length parameter is examined in Sect.~\ref{alpha-only}.
Furthermore, we suggest in Sect.~\ref{sec:metindex} that the parameter
can be employed as a metallicity indicator, with accuracies comparable
to other more widely used photometric indices. Our summary and
conclusions are given in Sect.~\ref{sec:sum-discussion}.

The nomenclature deserves a final note. We will be using several symbols for
our parameter throughout this paper. When talking about the parameter in a general
way, we will use the old symbol $W_{\rm HB}$, but when referring to measurements
specifically made for the $(B-V)$ or $(V-I)$ colors, we will use the
symbols $W^{B-V}_{\rm HB}$ and $W^{V-I}_{\rm HB}$, respectively.

\section{The data set \label{sec:datasets}}

For this paper we used two new, photometrically homogeneous databases that we
have recently created:

\begin{itemize}

\item The $V, I$ band ``ground-based'' dataset collected by 
Rosenberg et al (\cite{dutch00}, \cite{jkt00}) used for the GGC relative
age project by Rosenberg et al. (\cite{rel-ages99});

\item the ``HST snapshot'' dataset (Piotto et al. \cite{piottoEtal02}), based on F439W ($B$) and F555W ($V$) WFPC2
images of the core of all the GGCs with $(m-M)_B<18.0$, which has
already been used to constrain a number of parameters in the models
(Zoccali et al. \cite{zoccaliEtal00}, Piotto et al. \cite{piottoEtal00},
Bono et al. \cite{bonoEtal01}).

\end{itemize}

Both databases are available to the community on the Web pages of the Padova
Globular Cluster Group at http://menhir.pd.astro.it. 
 
For the present project, we used 
$26$ clusters in the
\noun{$B,\, V$} bands from the HST snapshot database, spanning a metallicity
interval  $-2.2<$[Fe/H]$<-0.5$, 
and $31$ clusters from the ground-based dataset, in the $V,\, I$ bands,
and covering a metallicity interval $-2.3<$[Fe/H]$<-1.1$.

The most important property of these databases for the aim of the
present project is their photometric internal homogeneity (cf.
Rosenberg et al. \cite{dutch00} and \cite{jkt00}, and Piotto et
al. \cite{piottoEtal02}). This homogeneity removes the main uncertainty in the
comparison done by B98. And, indeed, as exemplified in
Fig.~\ref{fig:m80vsm12}, the two new CMDs of M12 and M80 perfectly
overlap, giving the first confirmation that the spread noted by B98 was
mainly due to the photometric in-homogeneity of the data collected from
the literature.

\subsection{Measurement of the $W_{\rm HB}$  parameter  \label{sec:measurements}}

In order to compute the value of $W_{\rm HB}$, we first measured the
position of the HB turn-down for each cluster, and then the color of the
RGB at $0.5$ magnitudes above the HB-TD.

The HB-TD position is, using the definition of B98, the place where the
true color of the HB is zero. 
In order to find the TD on the observed diagrams, we first defined a
fiducial HB, that extends from the faint-blue tail to the bright-red
clump.
For the
\noun{hst} dataset,  NGC~2808 was used, since it already satisfies
our requirements (see e.g. Fig.~4 of Bedin et al. \cite{rolly00}).  On
the other hand, no cluster in the ground-based sample has such an
extended HB, so we constructed a mean fiducial HB, following the
procedure adopted in Rosenberg et al. (\cite{rel-ages99}). Briefly, we
started with the HB of NGC~1851, which has a bimodal HB, and then
extended it to the blue or the red by comparison with, respectively,
more metal poor and more metal rich clusters.

\begin{figure}
{\par\centering
\resizebox*{0.9\columnwidth}{!}{\includegraphics{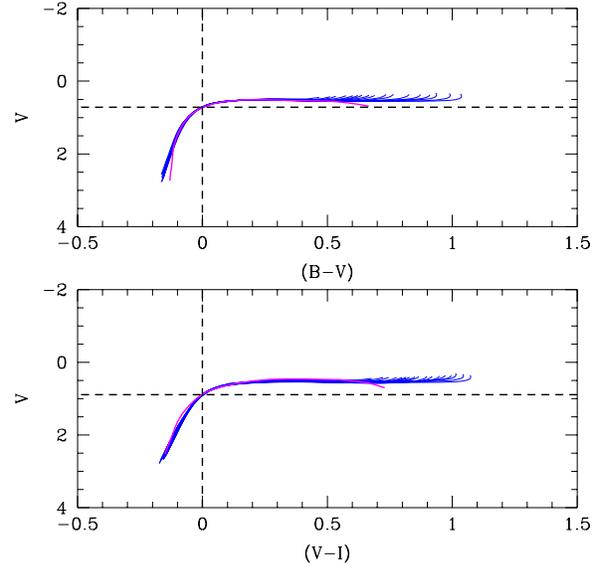}}
\par}
\caption{The TD position is shown on the V00 theoretical HBs in the case of the $(B-V)$
color (upper panel) and $(V-I)$ color (lower panel)\label{td}. 
Each line corresponds to the theoretical ZAHB sequence for a given
metallicity. Each one of them has been vertically shifted in this
plot, so that they all coincide at the TD point.
In the two panels, the empirical fiducial HBs have been represented as
well. They can be recognized as the solid curves ending at colors
$\approx 0.7$, and slightly fainter than the theoretical loci.
}
\end{figure}

\begin{figure}
{\par\centering \resizebox*{0.9\columnwidth}{!}{\includegraphics{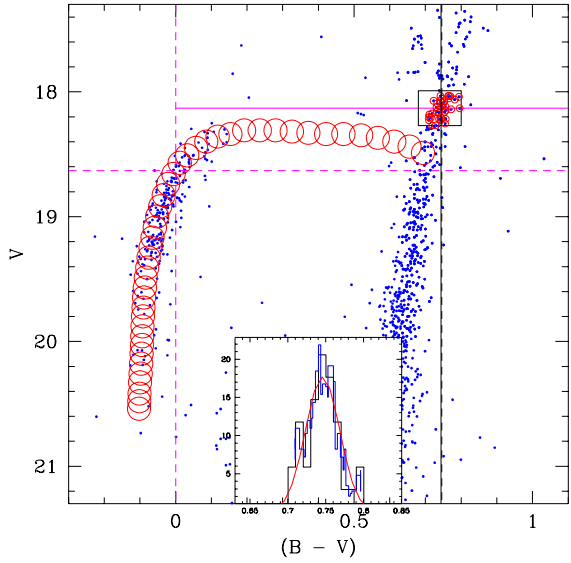}} \par}
\caption{An example of measurement of $W^{B-V}_{\rm HB}$, for NGC~5694.
This figure shows the mean fiducial HB in $(B-V),V$ (large open
circles), the position of the TD, and the corresponding color of the point on
the RGB which is $0.5$ magnitudes brighter than the TD.
The inset shows the histogram of the color distribution on the RGB, and the
solid curve is a Gaussian computed assuming the calculated mean and dispersion
of the data. \label{rgbbv}}
\end{figure}

\begin{figure}
{\par\centering \resizebox*{0.9\columnwidth}{!}{\includegraphics{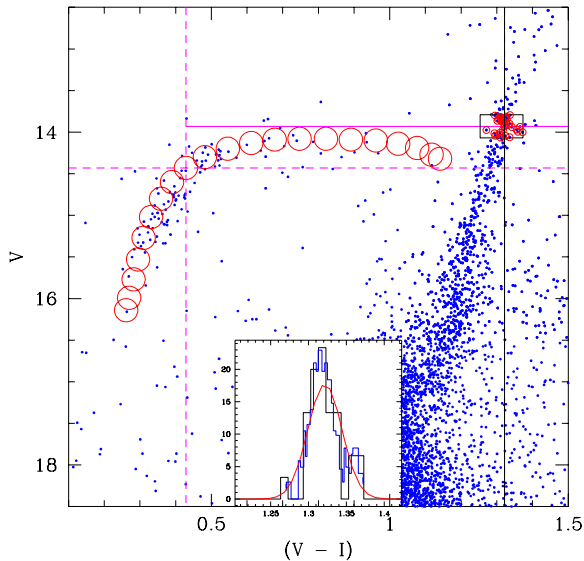}} \par}
\caption{Same as Fig.~\ref{rgbbv}, showing an example of measurement of $W^{V-I}_{\rm HB}$
for NGC~6656.\label{rgbvi}}
\end{figure}

The theoretical HBs from VandenBerg et al.\
(\cite{vandenberg00}; see Fig.~\ref{td}) were then fitted to the
fiducial HBs, and the TD-HB fixed at ($B-V$)=0.0, and ($V-I$)=0.0, as in
B98.  
For each cluster, we applied a color and a magnitude offset until
a satisfying visual superposition to the reference HB was obtained
(cf. Figs.~\ref{rgbbv},
\ref{rgbvi}). Finally the TD color and magnitude for each cluster are
those of the mean fiducial plus the two offsets.  
A special
case is that of NGC~6388 and NGC~6441, whose HBs are evidently sloped,
such that the blue portion just before the HB-TD is brighter than the
red portion (see Piotto et al. \cite{piottoEtal02}). There are enough
stars on the blue side, around the HB-TD, to allow a determination of
the HB-TD position, without relying on the red part of the HB. However,
if we tried to reach an overall agreement from blue to red, then a
slightly fainter HB-TD would have been found. In turn, this implies a
smaller $W_{\rm HB}$, and one can see in Fig.~\ref{fig:Wb} and the following,
that the representative points of the two cluster would move toward 
the general trend defined by the isochrones.

In order to measure the color of the RGB, we first obtained by hand an
approximate color of the RGB region $0.5$ magnitude brighter than the
TD level.  The final color was computed as the median color of all the
stars comprised in a rectangular region around this first RGB position
estimate. The box dimensions were slightly varied from cluster to
cluster, so that the whole RGB color extension was comprised. The
{}``vertical{}'' size was chosen as to ensure that a statistically
significant number of stars entered the computation. Typical values
are $0.06$ magnitudes in color and $0.12$ magnitudes in $V$.  The
Figs. \ref{rgbbv} and \ref{rgbvi} show the final step of the
procedure for NGC~5694 (HST data) and NGC 6656 (groundbased data).

\subsection{Correction for differential reddening \label{sec:diffredd-correction}}

Since $W_{\rm HB}$ represents a difference in color, and since the
two points are about $1$ magnitude apart, we cannot neglect the
effect of the wavelength dependence of the reddening. It is known
that, given a mean reddening $E_{B-V}$, the value of the
absorption $A_{V}=R_{V}\times E_{B-V}$ varies with the star color
(temperature). In the case of heavily reddened clusters, we must also
take into account the dependence of $R_{V}$ on the absolute
reddening $E_{B-V}$.

A first investigation on these effects was carried out by Olson
(\cite{Olso75}), who proposed a correction for the $(B-V)$
color. Later, Grebel and Roberts (\cite{greb95}; GR95)
extended this study by determining the trends of $R$, $A_{V}$,
colors, and reddenings, not only as a function of the temperature, but also
on star's gravity and metallicity. The results of GR95 are consistent
with Olson (\cite{Olso75}).

We used the results from both studies to correct the $W_{\rm HB}$ 
parameter measured in
the $B,\, V$ bands.
In order to correct the $W_{\rm HB}$ parameter obtained in the $V$
vs. $(V-I)$ CMD, we used the GR95 tables, thus ensuring the
homogeneity of the results. More details on our calculations are given
in Appendix~\ref{sec:maths}.

\subsection{Metallicity scales}

Most values of the iron abundance for the single clusters have been
taken from Rutledge et al. (\cite{rhs97}; RHS97). More explicitly, we
have taken the [Fe/H] values that RHS97 obtained by calibrating their
CaII triplet equivalent widths onto either the Zinn \& West
(\cite{zinnWest84}; ZW84) or the Carretta \& Gratton
(\cite{carrettaGratton97}; CG97) metallicity scales. For a few clusters,
there are no estimates of the metallicity provided by RHS97, and in such
cases the [Fe/H] values on the ZW84 scale are from Harris
(\cite{harris96}), and from Carretta (private communication) for the
CG97 scale. We have attached a typical error of $0.15$~dex to the
metallicities in this class.

\subsection{The $W_{\rm HB}$ parameter vs. metallicity \label{sec:thewhb-parameter}}

\begin{figure}
{\par\centering \resizebox*{1.0\columnwidth}{!}{\includegraphics{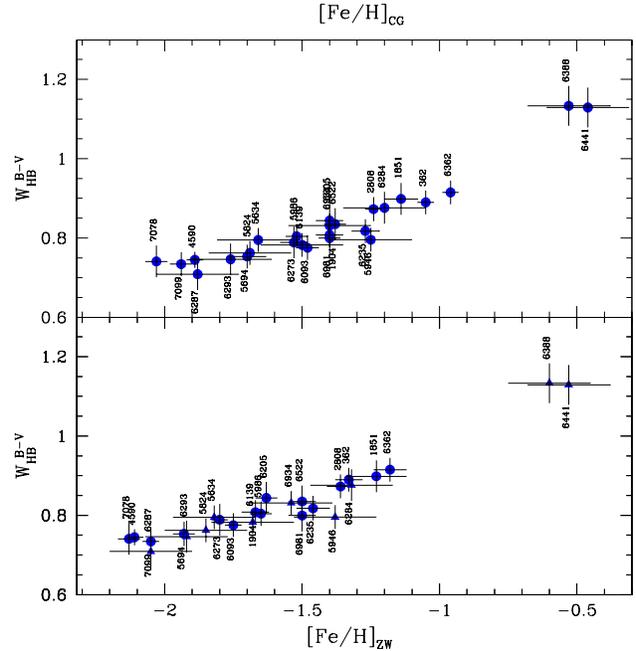}} \par}
\caption{All the data for $W^{B-V}_{\rm HB}$, as a function of
metallicity, adopting the CG97 ({\it top panel}) or ZW84 ({\it bottom
panel}) metallicity scales. Filled triangles identify clusters whose
metallicity on the ZW84 scale has been taken from Harris
(\cite{harris96}).
\label{fig_dati_bv}}
\end{figure}

\begin{figure}
{\par\centering \resizebox*{1.0\columnwidth}{!}{\includegraphics{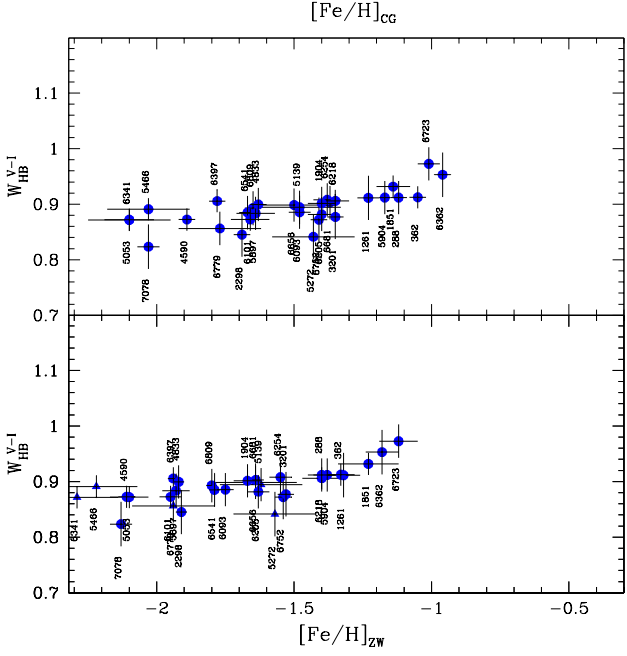}} \par}
\caption{The same as Fig.~\protect\ref{fig_dati_bv}, but for $W^{V-I}_{\rm HB}$.
\label{fig_dati_vi}}
\end{figure}

As in B98, Figs.~\ref{fig_dati_bv} and \ref{fig_dati_vi} display the
$W^{B-V}_{\rm HB}$ and $W^{V-I}_{\rm HB}$ parameters as a function of
{[}Fe/H{]}, both on the Zinn and West (\cite{zinnWest84}, {\it lower
panel}) and the Carretta and Gratton (\cite{carrettaGratton97}, {\it
upper panel}) metallicity scales. The same data is presented in
Table~1. The first remarkable result, at variance with B98,
is the low dispersion of the data points, fully compatible with the
error bars.  The dispersion at intermediate metallicities noted in B98,
is no longer present.  As originally suspected, the anomalous trend in
the B98 data was likely due to calibration errors present in their CMD
database, which was a simple collection of literature data.  Again, this
result shows the importance of using a photometric homogeneous database
in deriving the properties of the stellar population of star clusters.

\begin{table}
\caption{Measured values of $W_{\rm HB}$ for our sample. An underlined
$0.15$~dex error in the metallicity, highlights a value of the [Fe/H]
taken from Harris (\cite{harris96}). \label{tab_dati} }
\begin{tabular}{llll}
\hline\noalign{\smallskip}
Id. & [Fe/H]$_{\rm ZW}$ & [Fe/H]$_{\rm CG}$ & $W^{B-V}_{\rm HB}$ \\
\noalign{\smallskip}\hline\noalign{\smallskip}
NGC\,362  & $ -1.33 \pm 0.05 $  & $ -1.05 \pm 0.03 $  & $ 0.88 \pm 0.03 $  \\
NGC\,1851 & $ -1.23 \pm 0.11 $  & $ -1.14 \pm 0.06 $  & $ 0.89 \pm 0.04 $  \\
NGC\,1904 & $ -1.67 \pm 0.05 $  & $ -1.40 \pm 0.05 $  & $ 0.80 \pm 0.03 $  \\
NGC\,2808 & $ -1.36 \pm 0.05 $  & $ -1.24 \pm 0.03 $  & $ 0.87 \pm 0.03 $  \\
NGC\,4590 & $ -2.11 \pm 0.03 $  & $ -1.89 \pm 0.03 $  & $ 0.74 \pm 0.02 $  \\
NGC\,5634 & $ -1.82 \pm \underline{0.15} $  & $ -1.66 \pm\underline{0.15} $  & $ 0.79 \pm 0.03 $  \\
NGC\,5694 & $ -1.93 \pm 0.04 $  & $ -1.70 \pm 0.07 $  & $ 0.75 \pm 0.03 $  \\
NGC\,5824 & $ -1.85 \pm \underline{0.15} $  & $ -1.69 \pm\underline{0.15} $  & $ 0.76 \pm 0.03 $  \\
NGC\,5946 & $ -1.38 \pm \underline{0.15} $  & $ -1.25 \pm\underline{0.15} $  & $ 0.79 \pm 0.03 $  \\
NGC\,5986 & $ -1.65 \pm 0.04 $  & $ -1.52 \pm 0.04 $  & $ 0.80 \pm 0.03 $  \\
NGC\,6093 & $ -1.75 \pm 0.03 $  & $ -1.48 \pm 0.04 $  & $ 0.77 \pm 0.03 $  \\
NGC\,6139 & $ -1.68 \pm \underline{0.15} $  & $ -1.50 \pm\underline{0.15} $  & $ 0.78 \pm 0.03 $  \\
NGC\,6205 & $ -1.63 \pm 0.04 $  & $ -1.40 \pm 0.05 $  & $ 0.84 \pm 0.04 $  \\
NGC\,6235 & $ -1.46 \pm 0.06 $  & $ -1.27 \pm 0.05 $  & $ 0.81 \pm 0.03 $  \\
NGC\,6273 & $ -1.80 \pm 0.03 $  & $ -1.53 \pm 0.05 $  & $ 0.78 \pm 0.04 $  \\
NGC\,6284 & $ -1.32 \pm \underline{0.15} $  & $ -1.20 \pm\underline{0.15} $  & $ 0.87 \pm 0.04 $  \\
NGC\,6287 & $ -2.05 \pm \underline{0.15} $  & $ -1.88 \pm\underline{0.15} $  & $ 0.70 \pm 0.04 $  \\
NGC\,6293 & $ -1.92 \pm \underline{0.15} $  & $ -1.76 \pm\underline{0.15} $  & $ 0.74 \pm 0.04 $  \\
NGC\,6362 & $ -1.18 \pm 0.06 $  & $ -0.96 \pm 0.03 $  & $ 0.91 \pm 0.03 $  \\
NGC\,6388 & $ -0.60 \pm \underline{0.15} $  & $ -0.53 \pm\underline{0.15} $  & $ 1.13 \pm 0.05 $  \\
NGC\,6441 & $ -0.53 \pm \underline{0.15} $  & $ -0.46 \pm\underline{0.15} $  & $ 1.12 \pm 0.05 $  \\
NGC\,6522 & $ -1.50 \pm 0.05 $  & $ -1.38 \pm 0.04 $  & $ 0.83 \pm 0.04 $  \\
NGC\,6934 & $ -1.54 \pm \underline{0.15} $  & $ -1.40 \pm\underline{0.15} $  & $ 0.83 \pm 0.03 $  \\
NGC\,6981 & $ -1.50 \pm 0.05 $  & $ -1.40 \pm 0.04 $  & $ 0.79 \pm 0.04 $  \\
NGC\,7078 & $ -2.13 \pm 0.04 $  & $ -2.03 \pm 0.04 $  & $ 0.74 \pm 0.04 $  \\
NGC\,7099 & $ -2.05 \pm 0.03 $  & $ -1.94 \pm 0.04 $  & $ 0.73 \pm 0.03 $  \\
\noalign{\smallskip}\hline\noalign{\smallskip}
Id. & [Fe/H]$_{\rm ZW}$ & [Fe/H]$_{\rm CG}$ & $W^{V-I}_{\rm HB}$ \\
\noalign{\smallskip}\hline\noalign{\smallskip}
NGC\,288  & $ -1.40 \pm 0.05 $  & $ -1.12 \pm 0.03 $  & $ 0.91 \pm 0.03 $  \\
NGC\,362  & $ -1.33 \pm 0.05 $  & $ -1.05 \pm 0.03 $  & $ 0.91 \pm 0.02 $  \\
NGC\,1261 & $ -1.32 \pm 0.06 $  & $ -1.23 \pm 0.04 $  & $ 0.91 \pm 0.04 $  \\
NGC\,1851 & $ -1.23 \pm 0.11 $  & $ -1.14 \pm 0.06 $  & $ 0.93 \pm 0.02 $  \\
NGC\,1904 & $ -1.67 \pm 0.05 $  & $ -1.40 \pm 0.05 $  & $ 0.90 \pm 0.03 $  \\
NGC\,2298 & $ -1.91 \pm 0.02 $  & $ -1.69 \pm 0.03 $  & $ 0.84 \pm 0.04 $  \\
NGC\,3201 & $ -1.53 \pm 0.03 $  & $ -1.35 \pm 0.03 $  & $ 0.87 \pm 0.04 $  \\
NGC\,4590 & $ -2.11 \pm 0.03 $  & $ -1.89 \pm 0.03 $  & $ 0.87 \pm 0.02 $  \\
NGC\,4833 & $ -1.92 \pm 0.02 $  & $ -1.63 \pm 0.03 $  & $ 0.89 \pm 0.03 $  \\
NGC\,5053 & $ -2.10 \pm 0.07 $  & $ -2.10 \pm 0.09 $  & $ 0.87 \pm 0.02 $  \\
NGC\,5139 & $ -1.62 \pm \underline{0.15} $  & $ -1.48 \pm\underline{0.15} $  & $ 0.89 \pm 0.03 $  \\
NGC\,5272 & $ -1.57 \pm \underline{0.15} $  & $ -1.43 \pm\underline{0.15} $  & $ 0.84 \pm 0.04 $  \\
NGC\,5466 & $ -2.22 \pm \underline{0.15} $  & $ -2.03 \pm\underline{0.15} $  & $ 0.89 \pm 0.02 $  \\
NGC\,5897 & $ -1.93 \pm 0.05 $  & $ -1.64 \pm 0.07 $  & $ 0.88 \pm 0.03 $  \\
NGC\,5904 & $ -1.38 \pm 0.05 $  & $ -1.17 \pm 0.02 $  & $ 0.91 \pm 0.03 $  \\
NGC\,6093 & $ -1.75 \pm 0.03 $  & $ -1.48 \pm 0.04 $  & $ 0.88 \pm 0.03 $  \\
NGC\,6101 & $ -1.95 \pm 0.04 $  & $ -1.66 \pm 0.07 $  & $ 0.87 \pm 0.02 $  \\
NGC\,6205 & $ -1.63 \pm 0.04 $  & $ -1.40 \pm 0.05 $  & $ 0.88 \pm 0.03 $  \\
NGC\,6218 & $ -1.40 \pm 0.07 $  & $ -1.35 \pm 0.05 $  & $ 0.90 \pm 0.02 $  \\
NGC\,6254 & $ -1.55 \pm 0.04 $  & $ -1.38 \pm 0.05 $  & $ 0.90 \pm 0.03 $  \\
NGC\,6341 & $ -2.29 \pm \underline{0.15} $  & $ -2.10 \pm\underline{0.15} $  & $ 0.87 \pm 0.02 $  \\
NGC\,6362 & $ -1.18 \pm 0.06 $  & $ -0.96 \pm 0.03 $  & $ 0.95 \pm 0.04 $  \\
NGC\,6397 & $ -1.94 \pm 0.02 $  & $ -1.78 \pm 0.03 $  & $ 0.90 \pm 0.02 $  \\
NGC\,6541 & $ -1.79 \pm 0.02 $  & $ -1.67 \pm 0.03 $  & $ 0.88 \pm 0.03 $  \\
NGC\,6656 & $ -1.64 \pm \underline{0.15} $  & $ -1.50 \pm\underline{0.15} $  & $ 0.89 \pm 0.03 $  \\
NGC\,6681 & $ -1.64 \pm 0.03 $  & $ -1.37 \pm 0.03 $  & $ 0.90 \pm 0.03 $  \\
NGC\,6723 & $ -1.12 \pm 0.07 $  & $ -1.01 \pm 0.04 $  & $ 0.97 \pm 0.03 $  \\
NGC\,6752 & $ -1.54 \pm 0.03 $  & $ -1.41 \pm 0.03 $  & $ 0.87 \pm 0.04 $  \\
NGC\,6779 & $ -1.94 \pm \underline{0.15} $  & $ -1.77 \pm\underline{0.15} $  & $ 0.85 \pm 0.03 $  \\
NGC\,6809 & $ -1.80 \pm 0.02 $  & $ -1.65 \pm 0.03 $  & $ 0.89 \pm 0.03 $  \\
NGC\,7078 & $ -2.13 \pm 0.04 $  & $ -2.03 \pm 0.04 $  & $ 0.82 \pm 0.04 $  \\
\noalign{\smallskip}\hline
\end{tabular}
\end{table}

The linear dependence of $W_{\rm HB}$ on metallicity suggested by
B98 is not confirmed by Figs.~\ref{fig_dati_bv} and \ref{fig_dati_vi}:
there is a break in the linear trend at about [Fe/H]$\sim-1.4$. This
is more evident for $W^{B-V}_{\rm HB}$, thanks to the larger coverage in
metallicity of the HST data.

Figs.~\ref{fig_dati_bv} and \ref{fig_dati_vi}, shows also that a given
change in {[}Fe/H{]} produces a change in $W^{B-V}_{\rm HB}$ which
is $\sim 1.5$ times larger than that in $W^{V-I}_{\rm HB}$.

\section{Comparison with theoretical models \label{sec:compare-models}}

\begin{figure}
{\par\centering \resizebox*{1.0\columnwidth}{!}{\includegraphics{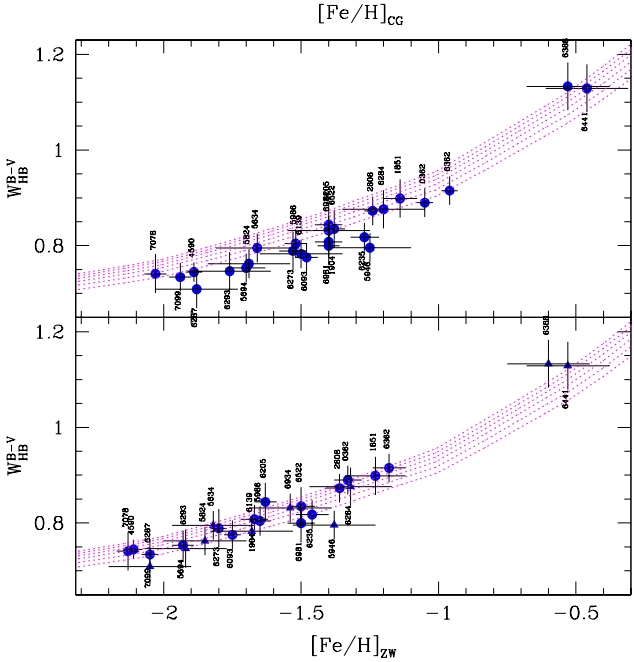}} \par}
\caption{$W_{\rm HB}^{(B-V)}$ as a function of metallicity, 
using the \noun{hst}
data sample. The CG97 scale is used in the top panel, while ZW84 is
used in the lower one. The dashed lines reproduce the
theoretical parameter calculated using the V00 isochrones, between
${8\textrm{ Gyr}}$ (lower line) and ${18\textrm{ Gyr}}$ ({\it upper
line}), and in $2$~Gyr steps. \label{fig:Wb}}
\end{figure}

\begin{figure}
{\par\centering \resizebox*{1.0\columnwidth}{!}{\includegraphics{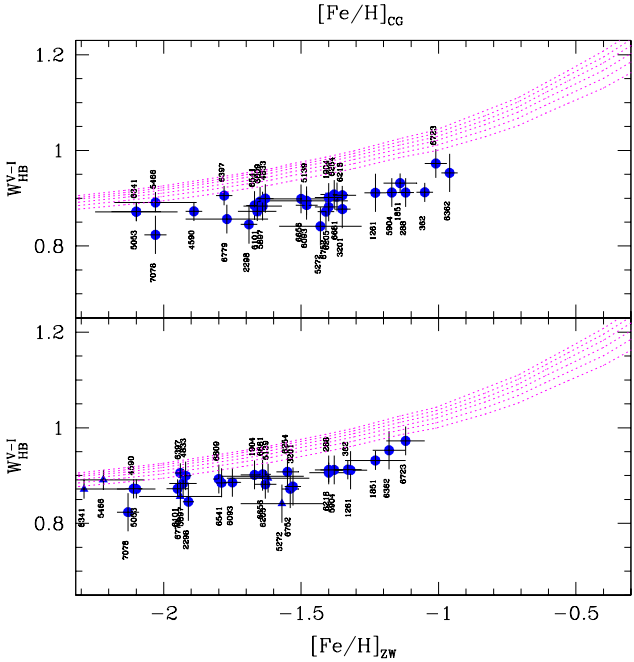}} \par}
\caption{Same as Fig.~\ref{fig:Wb}, for the ground-based dataset. 
\label{fig:Wi}}
\end{figure}

The interpretation of the data was carried out using both the VandenBerg
et al.  (\cite{vandenberg00}) and Girardi et
al. (\cite{girardiEtal00}) isochrones. 

We first present the two isochrone sets, and then show the theoretical
prediction for $W_{\rm HB}$ vs. [Fe/H] together with the empirical data
sets. The thorough inter-comparison is then carried out in
Sect.~\ref{sec:what-det-whb}.

\subsection{The VandenBerg et al.\ (\cite{vandenberg00}) isochrones}

The V00 isochrones are computed for $ 17 $ values of {[}Fe/H{]} ($
-2.3\leq \rm [Fe/H]\leq -0.3 $), $ 6 $ age values ($ 8\leq
t\leq 18 $ Gyr), and a fixed mixing length parameter $ \alpha_{\rm
MLT}=1.89 $. These models were computed for a fixed value of {[}$
\alpha $/Fe{]}$ =0.3 $, where $ \alpha $ stands for the alpha-elements
O, Ne, Na, Mg, Si, S, Ar, Ca and Ti.  The isochrones have been
{\changed transformed} to the observational plane by means of semi-empirical
color-\Teff\ relations (see VandenBerg et al. \cite{vandenberg00}) that
are essentially theoretical for $\Teff>5000$~K (where the TD is), and
empirically-corrected for lower \Teff\,s (where the RGBs are).

The V00 database includes smooth and well-behaved ZAHB sequences for
all their metallicity values. They are shown in Fig.~\ref{td}, for
both the $ B-V $ and the $ V-I $ colors. All ZAHB
sequences have been shifted vertically in this plot, so that they
coincide at the point where the color is 0, i.e.\ at the turn-down.
No color shift has been applied, further confirming that the HB-TD
point is fixed in color.  This figure has also been used to define the
``mean HB'' mentioned previously in Sect.~\ref{sec:measurements}.

The trends with metallicity of $ W_{\rm HB}^{B-V} $ and $ W_{\rm
HB}^{V-I} $ from V00 models (dashed lines) are compared with the data
in Figs.~\ref{fig:Wb} and \ref{fig:Wi}, respectively. As it can be noticed,
these models reproduce reasonably well the data for $ W_{\rm HB}^{B-V} $, 
especially if the ZW scale is adopted, but present $ W_{\rm HB}^{V-I} $ 
values that are systematically larger than the observed ones by about 
0.1~mag. 
This point is thoroughly discussed in Sect.~\ref{sec:what-det-whb}.
It is also clear that $ W_{\rm HB}$ increases with age, since
the Hayashi track becomes redder as the mass of evolved giants decrease
(see also Sect.~\ref{sec:what-age}). 

\subsection{The Girardi et al.\ (\cite{girardiEtal00}) isochrones 
\label{sec:isoc-g00}}

The G00 models were calculated for metallicities $ Z=0.0004 $, $0.001$,
$0.004$, $0.008$, $0.019$ and $0.03$ (Girardi et
al. \cite{girardiEtal00}). An additional set with $ Z=0.0001 $ (or
[Fe/H]$=-2.28$) has been computed 
adopting identical input physics (Girardi, unpublished). All models have
been computed with scaled-solar metal ratios and a helium content that
mildly increases with $ Z $, i.e.\ $ Y=0.23+2.75\, Z $. In this case,
the relation $ \feh =\log (Z/0.019) $ gives \feh\ values accurate to
within 0.03~dex. Thus, the 5 lowest $ Z $ values we are going to
consider (0.0001, 0.0004, 0.001, 0.004, and 0.008) would correspond to
\feh\ values of $ -2.28 $, $ -1.68 $, $ -1.28 $, $ -0.68 $, and $ -0.38
$. The
mixing length parameter was assumed to be $ \alpha _{\rm MLT}=1.68 $.

The Girardi et al. (\cite{girardiEtal00}) models are computed with
OPAL (Rogers \& Iglesias \cite{rogersIglesias92}) and Alexander \&
Ferguson (\cite{alexanderFerguson94}) opacities, and an updated
equation of state that considers the main non-ideal effects, such as
the Debye-Huckel correction. In these aspects, their lowest-mass
models (oldest isochrones) are very similar to many of the
non-diffusive stellar models available in the literature (see e.g.\
Weiss \& Schlattl \cite{weissSchalttl98}).  However, due to small
differences in the adopted physics of electron-degenerate matter, and
in the way used to generate ZAHB structures, Girardi et al.\ models
present ZAHB luminosities that are systematically lower (by $ \sim 0.2
$~mag) than other recent calculations (see Castellani et
al. \cite{castellaniEtal00}, and Salasnich 2001, for all details).  
These difference in luminosity seems
to be systematic and almost independent of metallicity: in fact, G00
ZAHB models obey the relation $ M_{V}^{\textrm{RRLyr}}=(0.163\pm0
.015)\, \feh +{\textrm{constant}.} $ (see Salasnich
\cite{salasnich01}), whose slope is consistent with those typically
found in most recent models of stellar evolution, i.e.\ $0.18~\rm mag/dex$
(e.g.\ Cassisi et al. \cite{cassisiEtal99}).

\begin{figure}
{\par\centering \resizebox*{1\columnwidth}{!}{\includegraphics{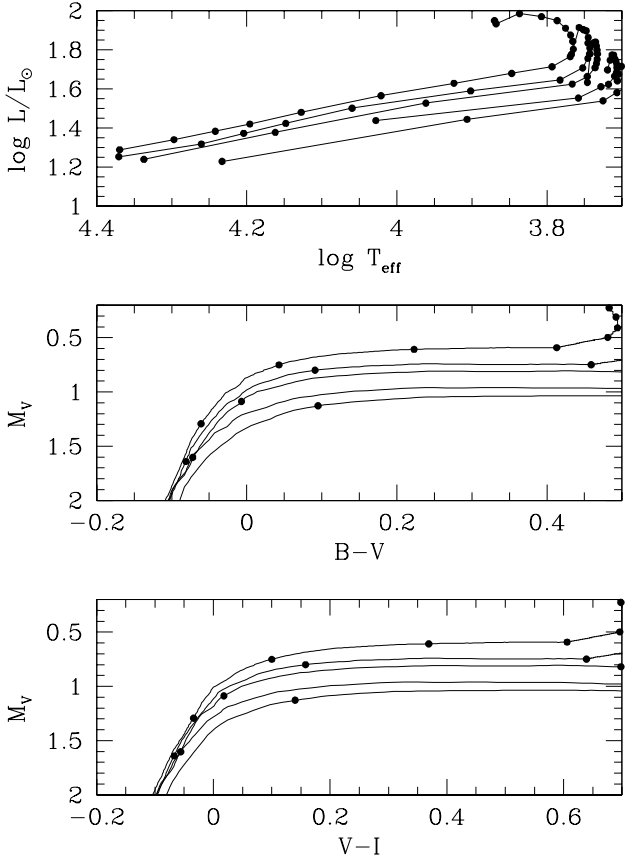}} \par}
\caption{ZAHB sequences from G00 models.}
\label{fig_zahb}
\end{figure}

The G00 core-helium burning models are quite complete and well-sampled
for masses higher than 0.6~\Msun. For lower masses, the ZAHB is not as
well covered as in V00 models and, as a consequence, the
identification of the HB-TD is not as straightforward. 
To this aim,
we proceeded in the
following way: First, we artificially obtained a more detailed ZAHB
sequence in the HR diagram, by simply interpolating between the ZAHB
models that have been actually calculated for just a few masses
$M<0.6$~\Msun.  This is displayed in the upper panel of
Fig.~\ref{fig_zahb}.  As it can be noticed, the computed models are
disposed along sequences that are fairly linear in the temperature
interval $ \log \teff >3.8 $. Thus, we can safely adopt linear
interpolations between these different models. Also the mass values
are linearly interpolated, which represents a less accurate, but not
critical, approximation.

Then, the interpolated ZAHB sequences are transformed to the
observational quantities by adopting the metallicity-dependent
bolometric corrections and color transformations from Bertelli et al.\
(\cite{bertelliEtal94}). In the \teff\ interval we are dealing with,
these transformations are entirely based on Kurucz' (\cite{kurukz92})
library of synthetic spectra. They are mainly a function of \Teff\,
and just marginally depend on the surface gravity $ g $. Thus, any
possible inadequacy in the interpolation of stellar masses 
would not be critical in this step.

The results for the \mv\ vs. \bv\ and \mv\ vs. \vi\ diagrams are shown
in the middle and lower panels of Fig.~\ref{fig_zahb},
respectively. Looking at this figure, one can notice that the ZAHB
sequences of lower metallicities are systematically shifted to lower
(brighter) magnitudes.

The position of the turn-down is largely determined by the behavior of
the $ V $-band bolometric corrections as a function of either \Teff\
or color.  We define the turn-down as the point of each ZAHB sequence
for which $ \bv =0 $ (in the case of \mv\ vs. \bv\ diagrams), or $ \vi
=0 $ (for \mv\ vs. \vi\ diagrams).  Since the zero-points of the
{}``theoretical{}'' photometry are such that $ \bv =\vi =0 $ for the
A0 dwarf Vega, for giants of $ \bv \sim0 $ the differences between
\bv\ and \vi\ are of just of a few hundredths of magnitude. Hence, in
practice, the two different definitions of the turn-down correspond to
almost (but not exactly) the same ZAHB star.

\begin{figure}
{\par\centering \resizebox*{0.9\columnwidth}{!}{\includegraphics{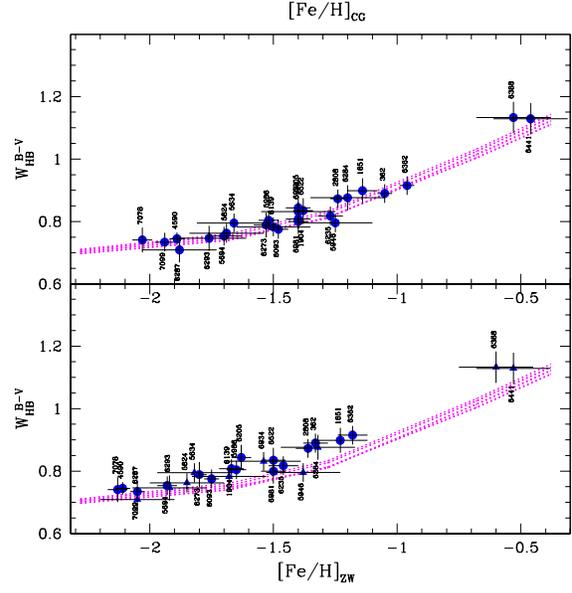}} \par}
\caption{The $W_{\rm HB}^{B-V}$ values, as measured from our
cluster sample (full dots with error bars, and marked with the
clusters' NGC number), are compared with the values derived from Girardi
et al. (\cite{girardiEtal00}) isochrones (dashed lines). Going from above to below,
isochrone ages are 16, 14, 12, and 10~Gyr.  \label{fig_wbv}}
\end{figure}

\begin{figure}
{\par\centering \resizebox*{0.9\columnwidth}{!}{\includegraphics{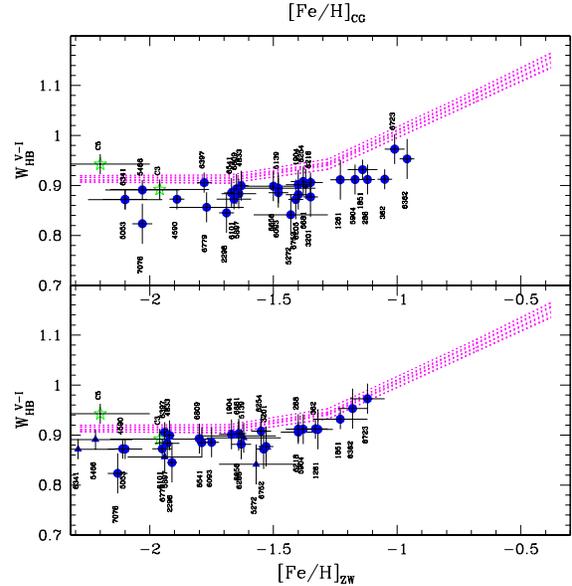}} \par}
\caption{The same as in Fig.~\ref{fig_wbv}, but for $ W_{\rm HB}^{V-I}$
\label{fig_wvi} }
\end{figure}

Having defined the ZAHB turn-down position for all metallicities, for
any isochrone we can measure $ W_{\rm HB}^{B-V} $ and $ W_{\rm
HB}^{V-I} $ by simply identifying the color of the point along the RGB
that is 0.5~mag brighter (in the $ V $-band) than its respective
turn-down.
The results for $ W_{\rm HB}^{B-V} $ are shown in Fig.~\ref{fig_wbv},
where the theoretical values obtained for isochrones with ages between
10 and 16 Gyr are compared to the observational data from
Sect.~\ref{sec:datasets}.  It can be noticed that $ W_{\rm HB}^{B-V} $
values increase steadily with metallicity, following the color 
shift of the RGB. If the Carretta \& Gratton
(\cite{carrettaGratton97}) metallicity scale is adopted, we find
a remarkable agreement between models and observations throughout the
complete metallicity range. The agreement becomes less satisfactory
if the data are plotted against the Zinn \& West (\cite{zinnWest84})
metallicity scale.

Similarly, the results for $ W_{\rm HB}^{V-I} $ are shown in
Fig.~\ref{fig_wvi}.  Again, $ W_{\rm HB}^{V-I} $ values are
found to increase steadily with metallicity in a way similar to the
one present in the data, though (i) in this case, there is no
data with $ \feh >-1 $ to compare the models with; and (ii) theoretical
$ W_{\rm HB}^{V-I} $ values are systematically larger than the data,
by about $ 0.05 $~mag .

\section{What determines $W_{\rm HB}$ \label{sec:what-det-whb} }

In this section, we will analyze the main parameters affecting the
theoretical value of $W_{\rm HB}$.

The four Figs.~\ref{fig:Wb}, \ref{fig:Wi},
\ref{fig_wbv}, and \ref{fig_wvi}, give us a chance to inter-compare
two recent sets of theoretical calculations with empirical data. It is
evident from the figures that both sets of isochrones reproduce
satisfactorily the trend of the observed points as a function of
[Fe/H].  The agreement is better for the $ BV $ data. 

For $ VI $, there is a disagreement of $\sim 0.05$~magnitudes between
the models and the data, which can be interpreted in two ways. Either
the models predict a smaller $ W_{\rm HB} $ value than observed, or
there is a problem with our measurements. Before proceeding further with
the discussion, we need to rule out the latter possibility. First, we
notice that the offset in color is independent of the cluster reddening,
so it cannot be attributed to our differential reddening corrections.
Then since $W_{\rm HB}$ is a differential quantity, the only
justification for such a spurious result would be the neglecting of the
quadratic (and higher order) terms in the calibration equations obtained
by Rosenberg et al. (\cite{dutch00}; \cite{jkt00}), such that a stretch of
$0.05$~magnitudes over a range of $1$~magnitude in color is produced.
However, the residual scatter around the assumed linear relations was a
few $0.001$ magnitudes over a range in color of more than $1.5$
magnitudes for the standard stars, so we must conclude that the problem
is indeed in the synthetic colors.

The main point that emerges from these figures, therefore, is that
theoretical models seem to describe reasonably well the relative
position of the HB-TD and RGB,
apart from a possible zero point difference in $V-I$.
Since the HB-TD color is constant, Figs.~\ref{fig:Wb}, \ref{fig:Wi},
\ref{fig_wbv}, and \ref{fig_wvi}
in fact represent the behavior of the RGB color, at about
the HB level, as a function of metallicity.

It is well-known that the RGB color in stellar models depends on a
series of factors.  First of all, it depends on metallicity.  Then, it
should strongly depend on the efficiency of the energy transport in
convective envelopes, and hence on the mixing length parameter $
\alpha $.  There must also be a weak dependence on the stellar mass,
and hence on the adopted isochrone age.  Some dependence on the model
chemistry (helium content and degree of $\alpha$-enhancement) may be
present.  Finally, the RGB color depends on the adopted
transformations between \Teff\ and color.  In the following, we will
explore these dependencies in order to verify whether the $W_{\rm HB}$
parameter is of some usefulness in setting one of the above
parameters.

\subsection{Dependence of $W_{\rm HB}$ on metallicity\label{sec:what-met}}

The dependence on the metallicity of $W_{\rm HB}$ is the most obvious 
and the strongest one. It is clearly visible in Figs.~\ref{rgbbv} and
\ref{rgbvi}. As the metallicity increases, the RGB becomes redder and
redder, while we do not expect a variation in the HB-TD color, and
therefore $W_{\rm HB}$ increases. The effect is stronger in ($B-V$)
than in ($V-I$). We will discuss this dependence in more details in Sect. 6.

\subsection{Dependence of $W_{\rm HB}$ on the mixing length 
parameter $\alpha$ \label{sec:what-mixlen}}

Among the various parameters fixing the RGB position, the mixing
length parameter $ \alpha $ plays one of the main roles, since in RGB
stars a significant fraction of the energy flux is transported by
convection.  The larger the value of $ \alpha$, the higher the
temperature of the RGB.

\begin{figure}
{\par\centering \resizebox*{1\columnwidth}{!}{\includegraphics{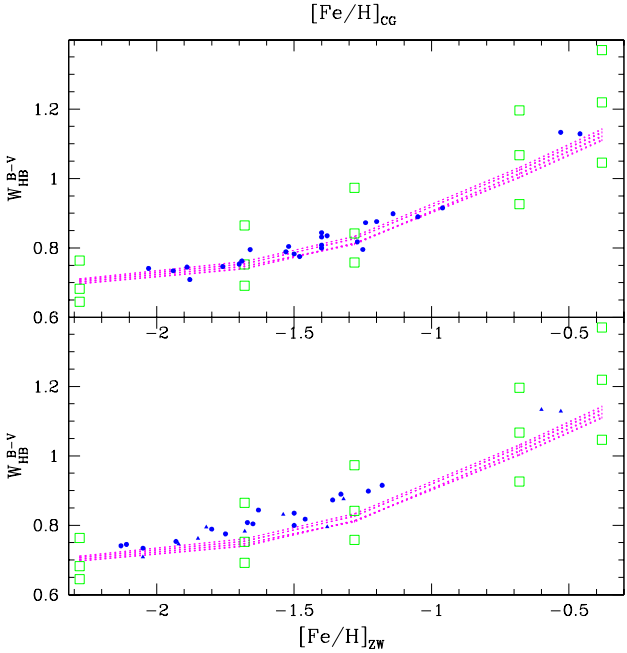}} \par}
\caption{Open squares: $ W_{\rm HB}^{B-V}$ values measured
for a set of 0.8~\Msun\ evolutionary tracks of varying metallicity and
mixing length parameter $ \alpha $. For each metallicity, $ \alpha $
values are 1.30, 1.68, and 2.00, going from above to below. For the sake
of comparison, we also plot the $ W_{\rm HB}^{B-V}$ values as derived
from the same isochrones as in Fig.~\protect\ref{fig_wbv} (dashed
lines), and the observational values for our cluster sample (small
dots). \label{fig_wbv_comp}}
\end{figure}

\begin{figure}
{\par\centering \resizebox*{1\columnwidth}{!}{\includegraphics{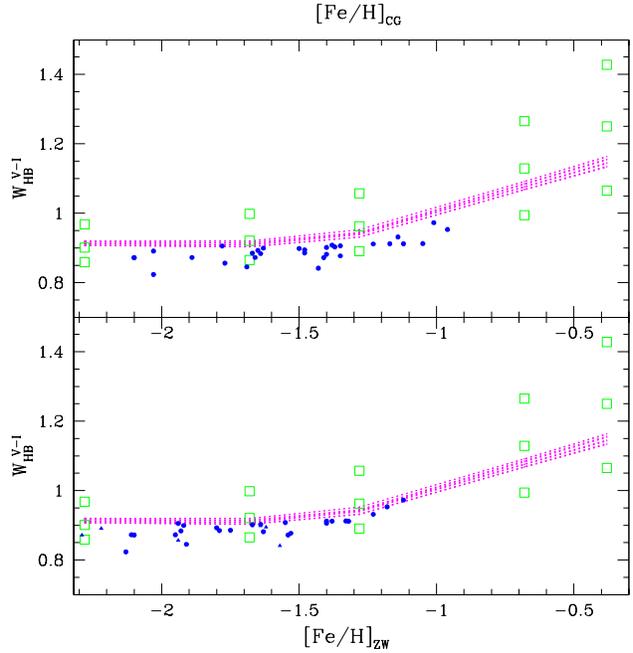}} \par}
\caption{The same as in Fig.~\protect\ref{fig_wbv_comp}, but for $ W_{\rm HB}^{V-I}$.
\label{fig_wvi_comp}}
\end{figure}

In order to evaluate the effect of $ \alpha $ in the $W_{\rm HB}$
parameter, we computed an additional set of stellar models, from the
zero age main sequence (ZAMS) up to the He-flash, for $ \alpha $ equal
to 1.30 and 2.00, respectively.  Only stars of 0.8~\Msun\ were
considered; this mass value is close the one found in the upper
RGB of isochrones that are 10 to 15~Gyr old. Together with the set of
$ \alpha =1.68 $ tracks already available (G00), the new tracks allow
us to evaluate how $ W_{\rm HB}^{B-V} $ and $ W_{\rm HB}^{V-I} $
change as a function of $ \alpha $. The $ W_{\rm HB}^{B-V} $ and $
W_{\rm HB}^{V-I} $ values for all our models with 0.8~\Msun, are
plotted in Figs.~\ref{fig_wbv_comp} and
\ref{fig_wvi_comp}, together with the values derived from our $ 10-16
$~Gyr old isochrones.

It can be noticed that, for lower metallicities, the points
corresponding to 0.8~\Msun, $ \alpha =1.68 $ models, coincide with the
locus of the isochrones.  For higher metallicities, this does not
happen, since metal-rich $ 10-16 $~Gyr isochrones typically present RGB
stars of higher masses ($ 0.9-1.0 $~\Msun), and hence hotter and bluer
than 0.8~\Msun\ RGB stars.

Other points to be noticed are that: (i) Models calculated with
different $ \alpha $ present virtually the same value of core mass at
the helium flash as the $ \alpha =1.68 $ ones, and a similar amount of
dredged-up helium.  This implies that the ZAHB models derived from these
tracks would present essentially the same luminosities. (ii) Stars
hotter than $ \Teff \simeq 7000$~K have their radii essentially
insensitive to the treatment of the outer convection, and therefore
independent of $ \alpha $ (see Renzini \& Fusi Pecci
\cite{renziniFusipecci88}; and figure 1 in Castellani et al. \cite{castellaniEtal99}).
Hence, even if we did not compute ZAHB models with different $ \alpha $
values, we know that the HB-TD position would not have changed.

Then, we can use the computed 0.8~\Msun\ RGB models to derive the
dependence of $W_{\rm HB}$ on $\alpha$: For each metallicity, we have
fitted straight lines to the $W_{\rm HB}^{B-V}$ versus $\alpha$, and
$W_{\rm HB}^{V-I}$ versus $\alpha$ data, using the 3 RGB tracks we have
for each metallicity. In all cases, the linear fit produced an
excellent description of the data. Therefore, we can conclude that
the $W_{\rm HB}^{B-V}(\alpha)$, and $W_{\rm HB}^{V-I}(\alpha)$
relations are very much linear.  The slopes of the fitted lines,
$\Delta W_{\rm HB}^{B-V}/\Delta \alpha$ and $\Delta W_{\rm
HB}^{V-I}/\Delta\alpha$, are presented in Table~\ref{tab_derivate}, for each
[Fe/H].

\begin{table*}
\caption{Derivatives of $ W_{\rm HB} $ cf.\ G00 models. \label{tab_derivate} }
\begin{tabular}{lllllll}
\hline\noalign{\smallskip}
 $Z$ & $\Delta W_{\rm HB}^{B-V}/\Delta\alpha$ & $\Delta W_{\rm
   HB}^{B-V}/\Delta({\rm age})$ & $\Delta(B-V)_{\rm RGB}$ & 
   $\Delta W_{\rm HB}^{V-I}/\Delta\alpha$ & 
   $\Delta W_{\rm HB}^{V-I}/\Delta({\rm age})$ & 
   $\Delta(V-I)_{\rm RGB}$ \\
 &  & $({\rm Gyr}^{-1})$ & (Alonso$-$Kurucz) &  & $({\rm Gyr}^{-1})$ & (Alonso$-$Kurucz) \\
\noalign{\smallskip}\hline\noalign{\smallskip}
 $0.0001$ & $-0.17$ & $0.0026$ & --      & $-0.16$ & $0.0020$ & --       \\
 $0.0004$ & $-0.25$ & $0.0036$ & --      & $-0.19$ & $0.0025$ & $-0.046$ \\
 $0.001 $ & $-0.31$ & $0.0038$ & --      & $-0.24$ & $0.0018$ & $-0.050$ \\
 $0.004 $ & $-0.38$ & $0.0053$ & $0.175$ & $-0.39$ & $0.0040$ & $-0.039$ \\
 $0.008 $ & $-0.46$ & $0.0057$ & $0.105$ & $-0.52$ & $0.0045$ & $-0.033$ \\
\noalign{\smallskip}\hline
\end{tabular}
\end{table*}

The numbers in Table~\ref{tab_derivate} confirm the strong dependence
of $W_{\rm HB}$ on $\alpha$, to be compared with the much lower
dependence on age.

\subsection{Dependence of $W_{\rm HB}$ on age \label{sec:what-age}}

The dependence of $W_{\rm HB}$ on age is already illustrated in 
Figs.~\ref{fig:Wb}, \ref{fig:Wi}, \ref{fig_wbv}, and \ref{fig_wvi}, 
where isocrohones for a wide range of ages are presented.
In Table~\ref{tab_derivate}, instead, we tabulate the derivatives of
$W_{\rm HB}$ with respect to age, at 14 Gyr, as derived from G00 models.
Surprisingly, in V00 models these derivatives are about twice as large; 
but anyway, it is clear that 
$W_{\rm HB}$ changes very little with age. Taking into account the small
age dispersion found by Rosenberg et al. (\cite{rel-ages99}) for the
same clusters used in the present investigation, we conclude that the
effects of age variations from cluster to cluster are much smaller than
the error bars of the single data points.

\subsection{Dependence of $W_{\rm HB}$ on the helium content \label{sec:what-helium}}

When the helium content is increased, the HB luminosity also increases
according to the law $ \Delta V/\Delta Y=-3.22 $ for $
0.23<Y<0.27 $, i.e.  the values suggested by current observations ($ R
$ parameter, HII regions, etc.). This means that, when measuring $
W_{\rm HB} $, we will consider a brighter RGB point as well. This
point will be redder, whereas the HB-TD does not change in color, so the
net result is that increasing $ Y $ will lead to a larger $ W_{\rm HB}
$. The size of this effect was measured using a $ 12 $~Gyr isochrone
from V00, varying {[}Fe/H{]} in the whole range, and for five values
of $ Y $.

\begin{figure}
{\par\centering \resizebox*{0.9\columnwidth}{!}{\includegraphics{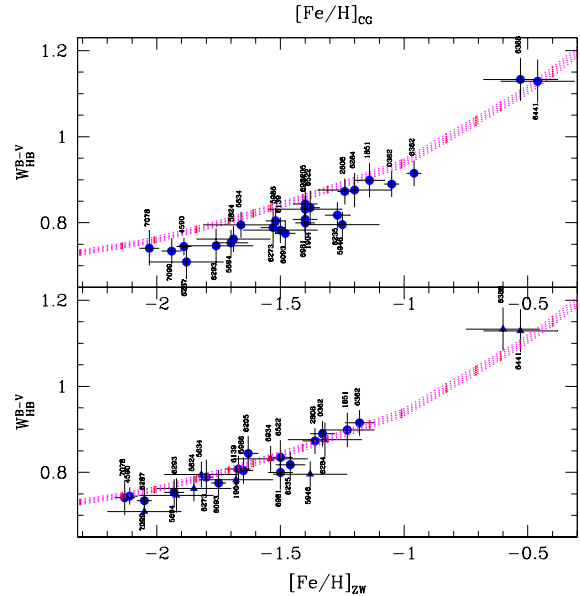}} \par}
\caption{The dashed lines show the effect, on $ W^{B-V}_{\rm HB}$,
of varying the helium content on the theoretical models. See text for
the explanation.
\label{fig:hebv}}
\end{figure}

\begin{figure}
{\par\centering \resizebox*{0.9\columnwidth}{!}{\includegraphics{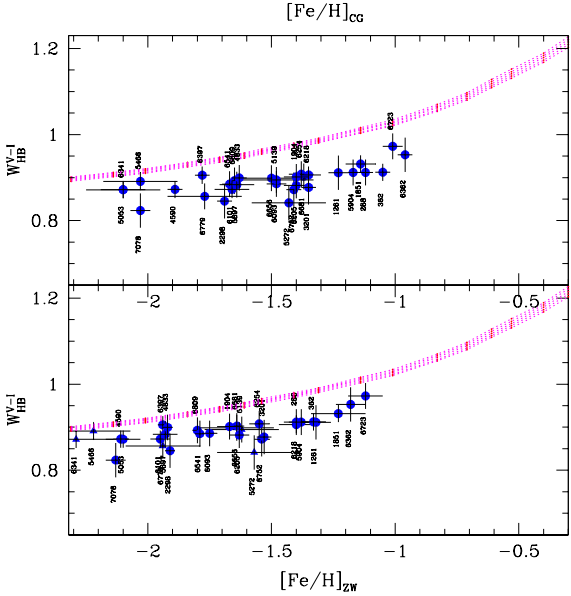}} \par}
\caption{Same as Fig.~\ref{fig:hebv}, for the ground-based observations. \label{fig:hevi}}
\end{figure}

The comparison with the observations is shown in Figs.~\ref{fig:hebv} and
\ref{fig:hevi}, for the \noun{hst} and ground-based samples, respectively.

The figures show that, while there is a zero-point problem (which could
be solved changing the reference age), the relative trend is well
reproduced by the models. In any case, variations in helium content have
a negligible effect on $ W_{\rm HB} $.
The dispersion of the empirical points due to the
observational uncertainties is larger than the dispersion we would
expect from a reasonable cluster to cluster variations in the helium
content.  Therefore, we conclude that the dependence of $ W_{\rm HB} $
on the He mass fraction can be ignored.

\subsection{Dependence of $W_{\rm HB}$ on $ \alpha $-enhanced metal ratios
\label{sec:what-enha}}

Halo populations
are expected to present $ \alpha $-enhanced metal ratios, with 
$[\alpha /{\textrm{Fe}}]\simeq0.3 $~dex.  For a given metal fraction $
 Z $, $ \alpha $-enhanced models have an iron content \feh\ that is
\emph{depleted} by a factor that corresponds roughly to the degree of
enhancement $ [\alpha /{\textrm{Fe}}] $. Moreover, for sufficiently
low metallicities, $ \alpha $-enhanced evolutionary tracks can be
replaced by their scaled-solar counterparts of same $ Z $ (see Salaris
et al. \cite{salarisEtal93}; Salaris \& Weiss \cite{salarisWeiss98}).

Therefore, for low metallicities we can expect that models with
solar-scaled abundances could be used to reproduce $\alpha$-enhanced
isochrones, provided that their \feh\ values are changed by about
$-0.3$~dex. Of course, this is only a first-order approximation to the
question, especially for the models of higher metallicities
($\feh\ga0.004$, see Salasnich et al. \cite{salasnichEtal00}). Another
approximation is in the fact that the \Teff-color transformations in
use have been derived from model atmospheres of scaled-solar
composition, and not from $\alpha$-enhanced ones (which are not yet
available). Whether this is critical, is yet to be investigated.

Keeping these points in mind, we can now get some insight on how much
$W_{\rm HB}$ depends on the $\alpha$-enhancement, by taking advantage
of the fact that V00 isochrones are enhanced while those of G00 are not.
As a caveat, we must stress that ideally one would like to measure the
size of the effect on otherwise identical models, since e.g. different
color transformations could mask the true trend. As a first order
approach, we will nevertheless try to understand what happens if we
interpret the differences in the two isochrone sets purely in terms of
$\alpha$-enhancement (i.e. as just an offset along the abscissae).

Let us then compare Fig.~\ref{fig:Wb} to Fig.~\ref{fig_wbv}, and 
\ref{fig:Wi} to \ref{fig_wvi}, looking for example at their [Fe/H]
position at $W_{\rm HB}=1$. We can see that $\alpha$-enhanced isochrones
look indeed more iron poor than solar-scaled isochrones. In particular,
the displacement for the $(B-V)$ color is $\Delta \rm [Fe/H] \simeq -0.1
$, while it is $\Delta \rm [Fe/H] \simeq -0.2$ for the $(V-I)$ color. The
effect then goes in the right direction, although the size of the offset
in the iron scale is less than expected (but remember the above
caveat).

If we now introduce the data, we see that the $(V-I)$ color is the one
that gives more troubles. There is a certain degree of agreement if
we consider the G00 theoretical trend, when data are plotted on the ZW84
metallicity scale. In all the other cases the isochrones look too iron
poor, regardless of the metallicity scale.

The case of the $(B-V)$ color is less clear-cut, since the choice of the
metallicity scale now plays an evident role. If we assume that the CG97
scale is correct, then the G00 models show the best agreement, so it
looks like no $\alpha$-enhancement is required. On the contrary, the V00
models are the ones that better reproduce the observed trend on the ZW84
metallicity scale. The G00 models should then be enhanced in order to
reach the same agreement.

In conclusion, both theoretical calculations show problems in
reproducing the $(V-I)$ trend of $W_{\rm HB}$ vs. [Fe/H], most probably
due to the color-$T_{\rm eff}$ relations (see also
Sect.~\ref{sec:what-transf}). If we then just rely on the $(B-V)$ color,
then it is really a matter of choosing one's preferred metallicity
scale. Since $\alpha$ elements are indeed enhanced in GGCs, and trusting
the models, one should choose the ZW84 scale and V00 isochrones. It is
also likely that G00 isochrones will be consistent with the data on the
ZW84 scale, once the $\alpha$-enhancement is introduced.  However, it
looks like the real way out should be a third independent determination
of the GGC metallicity scale. 

If the ZW84 scale could be demonstrated as the correct one, then there
would not be any need for a revision of the present picture (besides
some revision of the $(V-I) \div T_{\rm eff}$ relations). If instead the
CG97 scale is the good one, and retaining the $\alpha$-enhancement
scenario, then even $(B-V)$ color-temperature relations should be
revised.

\subsection{Dependence of $W_{\rm HB}$ on the \Teff-color transformations
\label{sec:what-transf}}

We have already noticed that the models that so well fit the $ W_{\rm
HB}^{B-V} $ data (Fig.~{\ref{fig_wbv}}), do not fit as well the $
W_{\rm HB}^{V-I} $ (Fig.~{\ref{fig_wvi}}) ones.  
In particular, while the theoretical trend seems to be the same as
the observed one, there is a zero point shift for
$ W_{\rm HB}^{V-I} $.
Since the isochrones we are using in these plots are the same, this is
probably indicating that we have problems with the color
transformations, either for \bv\ or \vi, or both.  In fact,
differences between Kurucz and empirical
\teff\ vs. color transformations have been noticed by several authors
(e.g.\ Gratton et al. \cite{grattonetal96}; Castelli et
al. \cite{castellietal97}; Weiss \& Salaris \cite{weissSalaris99}).
Notice that, in our isochrones, it would be enough to have the RGB just
0.05~mag 
bluer
in \vi\ to find a perfect agreement between models and
observations. A color shift of this order can be caused even by the
different filter transmission curves used by several authors.

Just to give an order of magnitude to the uncertainty due to the color
transformations, in Table~\ref{tab_derivate} we present the changes in
color, $\Delta(B-V)_{\rm RGB}$ and $\Delta(V-I)_{\rm RGB}$, that we
obtain for the RGB point 0.5~mag above the TD level in G00 12 Gyr-old
isochrones, in the case we adopt Alonso et al. (\cite{alonsoEtal99})
\Teff-color transformations instead of Kurucz (\cite{kurukz92})
ones. Unfortunately, such a difference can not be found for all values
of \feh, simply because at low metallicities we go out of the
applicability range of Alonso et al. (\cite{alonsoEtal99}) formulas (see
their tables 2 and 3).

As can be noticed, for higher metallicities we find consistent shifts
($\sim0.15$~mag) in the $B-V$ color of the RGB; however, it is not
possible to establish how these differences depend on metallicity, since
only 2 values were derived (for $Z=0.004$ and $0.008$). For $V-I$, instead,
the differences are surprisingly small ($\sim-0.04$~mag), and {\em seem not
to depend on metallicity}.

This exercise indicates that the uncertainties in \Teff-color 
transformations could sensibly affect our results for $W_{\rm HB}^{B-V}$,
but not for $W_{\rm HB}^{V-I}$. However, we remark that a comparison is
essentially missing for low metallicities. 

We note that Alonso et al. (\cite{alonsoEtal99}) constitutes the most 
updated source of empirical $\Teff$--color relations for
red giants. The comparison with other different transformations 
from the literature would not help much to
clarify the issue of the behavior of $W_{\rm HB}$.

\section{Can $W_{\rm HB}$ be used to calibrate the mixing length parameter
$\alpha$\,? \label{alpha-only}}

In the previous section  we showed that the mixing length parameter
$\alpha$ plays a major role in determining the values of $W_{\rm HB}$,
being by far more influent than the cluster age and helium content.

Moreover, a rapid inspection of Figs.~\ref{fig_wbv_comp} and
\ref{fig_wvi_comp} suggests that present data are confined within a
relatively narrow range of $\alpha$ values. A natural question that
comes to us, then, is: \emph{Is the data compatible with a single
value of the mixing length parameter $ \alpha$\,?}  And, if yes,
\emph{can we then use the data to single out a ``best value'' for
$\alpha$\,?}

In answering these questions, however, we should keep in mind 
the uncertainties in the \Teff--color transformations, 
and the enhancement of $\alpha$-elements, that are the main 
uncertainties in the comparison between models and data.
They have been shortly mentioned in the previous section.
Regarding the color transformations, we notice a mismatch 
in \vi\ data (Fig.~\ref{fig_wvi_comp}), and the possibility that
similar effects are present 
also in the \bv\ color. Considering this, 
we conclude that a single {}``best fitting value{}'' 
for $ \alpha $ cannot be derived, unless we know the
color transformations with an accuracy of some hundredths of
magnitude. The best we can do, for the moment,
is to check whether present models deviate
from the assumption of a single $ \alpha $ being valid throughout the
entire \feh\ range of our observations.
In this respect, the most important conclusion of 
Sect.~\ref{sec:what-transf} is that the offset introduced by changing
the color transformations seems not to depend on metallicity.

In order to 
make this check, we first measured the distance of each data data
point from a given set of isochrones (for a fixed age and $\alpha$),
$\Delta W_{\rm HB}$.  Then, we translated the $\Delta W_{\rm HB}$
values into differences in alpha, $\Delta \alpha$, using the $\Delta
W_{\rm HB}/\Delta \alpha$ derivative appropriate to each metallicity
(Table~\ref{tab_derivate}).  Then, any trend of $\Delta \alpha$ with
metallicity can be interpreted in terms of changes in the mixing
length parameter, under the assumption that no other parameter is
playing a significant role, as we have demonstrated in the previous
section.

\begin{figure}
{\par\centering \resizebox*{1\columnwidth}{!}{\includegraphics{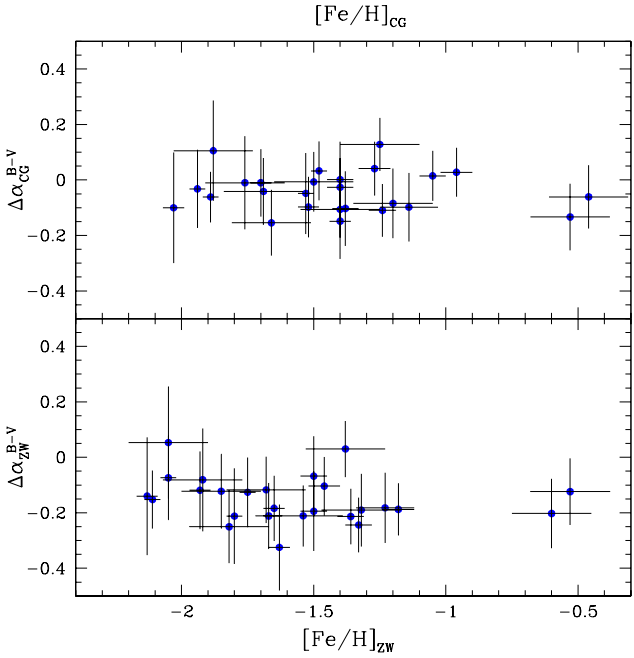}} \par}
\caption{The estimated $\Delta \alpha$ differences for each object
observed in $B-V$, and the reference G00 isochrone of age 14 Gyr, 
as a function of metallicity in the ZW ({\it bottom panel}) and 
CG97 scales ({\it upper panel}). Notice the
absence of any significant trend with metallicity.  \label{fig_alpha_bv}}
\end{figure}

\begin{figure}
{\par\centering \resizebox*{1\columnwidth}{!}{\includegraphics{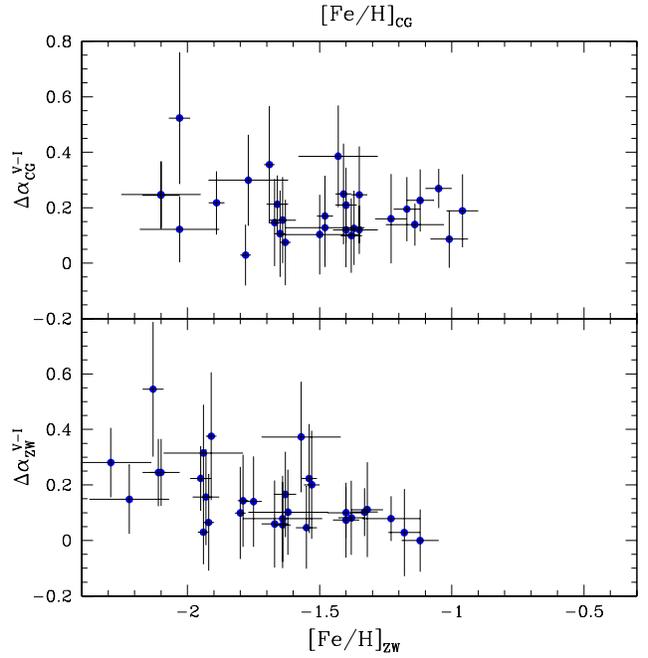}} \par}
\caption{The same as in Fig.~\protect\ref{fig_alpha_bv}, but for 
the clusters observed in $V-I$.
\label{fig_alpha_vi}}
\end{figure}

This was performed separately for $\Delta W_{\rm HB}^{B-V}$ and
$\Delta W_{\rm HB}^{V-I}$. Figs.~\ref{fig_alpha_bv} and
\ref{fig_alpha_vi} show the $\Delta \alpha$ values as a function of
metallicity, obtained having, as the reference, the G00 isochrones of
age 14~Gyr. Noticeably enough, the differences in $\alpha$ show no
significant trend with metallicity. The only case in which some
marginal trend of this kind could be present, is the one for $V-I$
data on the ZW scale, for which $\Delta W_{\rm HB}^{V-I}$ seems to
slightly decrease for $\feh>-1.5$.

We have then modeled the data in Figs.~\ref{fig_alpha_bv} and 
\ref{fig_alpha_vi} by least-squares fitting of either (i) a constant
value, or (ii) a straight line. The results are presented in
Table~\ref{tab_fit}.

\begin{table*}
\caption{Fitting parameters for the data in Figs.~\protect\ref{fig_alpha_bv} and 
\protect\ref{fig_alpha_vi}. \label{tab_fit} }
\begin{tabular}{llllll}
\hline\noalign{\smallskip}
\multicolumn{6}{l}{(i) A constant value: $\Delta\alpha = a$} \\
Case & $a$ & $\sigma_a$ & & & $\chi^2$ \\ 
\noalign{\smallskip}\hline\noalign{\smallskip}
CG scale, $BV$ data & $-$0.036 & 0.022 & & & 10.28 \\ 
ZW scale, $BV$ data & $-$0.157 & 0.025 & & & 9.82 \\ 
CG scale, $VI$ data & 0.178 & 0.022 & & & 13.11 \\ 
ZW scale, $VI$ data & 0.133 & 0.024 & & & 15.59 \\ 
\noalign{\smallskip}\hline\noalign{\smallskip}
\multicolumn{6}{l}{(i) A straight line: $\Delta\alpha = a + b\,({\rm [Fe/H]}+1.5)$} \\
Case & $a$ & $\sigma_a$ & $b$ & $\sigma_b$ & $\chi^2$ \\ 
\noalign{\smallskip}\hline\noalign{\smallskip}
CG scale, $BV$ data & $-$0.036 & 0.023 & $-$0.009 & 0.059 & 10.26 \\ 
ZW scale, $BV$ data & $-$0.157 & 0.025 & $-$0.027 & 0.064 & 9.64 \\ 
CG scale, $VI$ data & 0.174 & 0.024 & $-$0.028 & 0.066 & 12.92 \\ 
ZW scale, $VI$ data & 0.108 & 0.026 & $-$0.173 & 0.071 & 9.71 \\ 
\noalign{\smallskip}\hline
\end{tabular}
\end{table*}

In the case of $BV$ data, the fitted lines turned
out to have a negligible slope, and there is virtually no improvement
in the fittings as we pass from a constant to a straight line.  The
same applies for the $VI$ data in the CG97 scale.  The case of $VI$ data
on the ZW scale, with its mild slope ($-0.17\pm0.07$), constitutes the
only exception.

Perhaps more importantly is the fact that, assuming that $\alpha$ is
constant (first part of Table~\ref{tab_fit}), we get \emph{extremely
small} values for its dispersion, i.e. $\sigma_\alpha\simeq0.023$. This has
to be compared to typical values of $\alpha$ assumed in evolutionary
calculations, i.e.\ $\alpha\sim1.7$.  Overall, these numbers indicate
that, indeed, \emph{the empirical data can be very well approximated
by a constant value of $\alpha$.}

\section{A new metallicity index? \label{sec:metindex}}

\begin{figure}
{\par\centering \resizebox*{0.9\columnwidth}{!}{\includegraphics{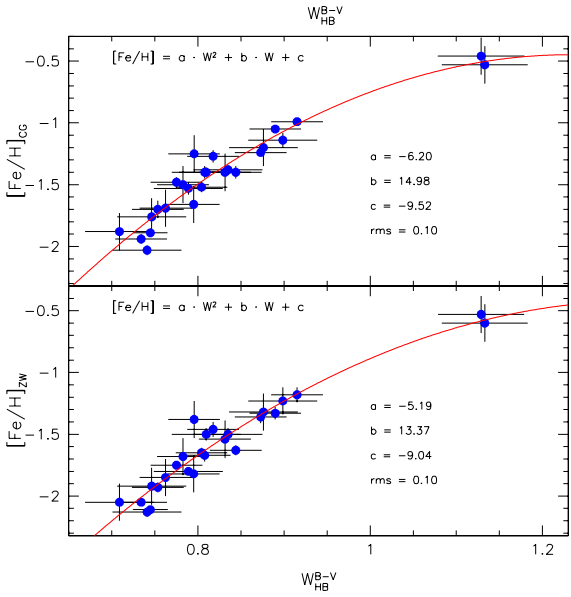}} \par}

\caption{Calibration of $ W_{\rm HB}^{B-V}$ as a metallicity index.
The observed data points have been interpolated with a second order polynomial.
\label{fig:parabola}}
\end{figure}

We have shown that there is a relation between $ W_{\rm HB} $ and the
metal content of the cluster. Since $ W_{\rm HB} $ has only a second
order dependence on the reddening (which can be corrected as described
in Appendix), it is a potentially interesting metallicity index.  We
have verified whether it is possible to use this parameter as a new
metallicity index. In Fig.~\ref{fig:parabola} we fitted a parabola to
the observed $W_{\rm HB}^{(B-V)}$ vs [Fe/H] relation for the
\noun{hst} sample, obtaining a $ 0.1 $~dex rms residual.

The sensitivity of the parameter, although lower than that of other
traditional indices, can be used to obtain a first estimate of the
metallicity. A typical $ 0.04 $~magnitude error on $ W^{B-V}_{\rm HB}$ 
would translate into a $ \sim 0.2 $~dex uncertainty on
{[}Fe/H{]}. On the other hand, the lack of metal-rich clusters,
and the much shallower dependence on metallicity of this parameter,
prevents a reliable calibration of $ W^{V-I}_{\rm HB} $.

\begin{figure}
{\par\centering \resizebox*{0.9\columnwidth}{!}{\includegraphics{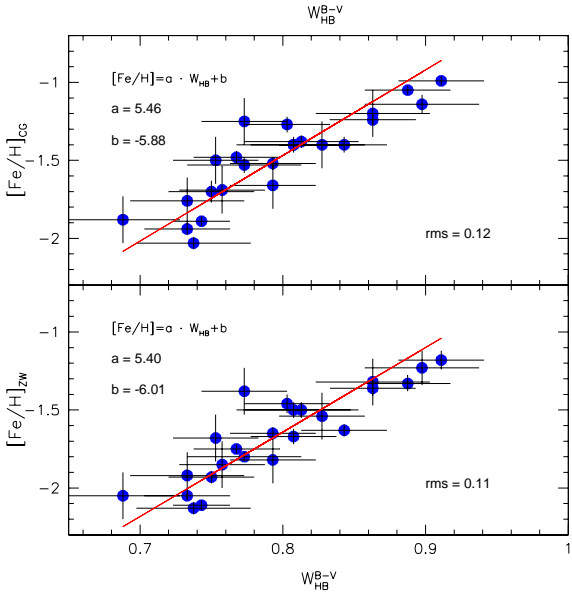}} \par}

\caption{Calibration of $ W^{B-V}_{\rm HB}$ for the \noun{hst}
clusters, obtained with the exclusion of NGC~6388 and NGC~6441. The parameters
of the weighted least square interpolation are reported in the figures, both
for the CG97 (upper panel) and ZW84 (lower panel) metallicity scales.\label{fig:rettabv}}
\end{figure}

\begin{figure}
{\par\centering \resizebox*{0.9\columnwidth}{!}{\includegraphics{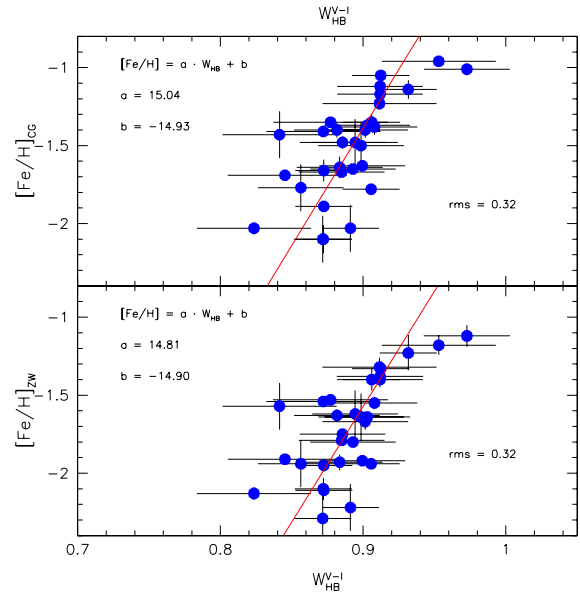}} \par}

\caption{Same as Fig.~\ref{fig:rettabv}, for the ground-based data. \label{fig:rettavi}}
\end{figure}

There could be some concern about the inclusion of the two metal-rich
clusters (NGC~6388 and NGC~6441) in the $ B-V $ calibration, since it
is still not known what drives the peculiar morphology of the HB of
these two objects. Therefore, we have also calibrated the $ W^{B-V}_{\rm HB}$ 
index in the $-2.1<$[Fe/H]$<-1.0$ metallicity interval.
In this case we used the weighted linear least squares fit, which is
shown in Figs.~\ref{fig:rettabv} and
\ref{fig:rettavi}. The result of the fits shows that the relation for
$ W^{B-V}_{\rm HB} $ has still a formally low dispersion ($ 0.1 $
dex rms), whereas a large $ 0.3 $~dex dispersion is shown by the residuals
for the $ W^{V-I}_{\rm HB} $ parameter.  Again, this confirms the higher
sensitivity of $ W^{B-V}_{\rm HB} $ to metallicity changes.

In Fig.~\ref{fig:and}, which shows the trend of the second order
reddening corrections that must be applied to $ W_{\rm HB} $, the
right vertical axis shows how a difference in $ W_{\rm HB} $
translates into a difference in metallicity, using the above
relations. It clearly shows that, especially for the highly reddened
clusters, neglecting the differential reddening corrections would lead
to significant systematic errors in the estimated metallicity. For
instance, the error on {[}Fe/H{]} for NGC~6287 would be of the order
of $ 0.2 $~dex.  Naturally, these errors are larger for metallicities
obtained using the $ V-I $ calibration.

In any case, we must note that: (i) The rms error are just formal
fitting errors, while the real uncertainty on the metallicity is
higher, as it must take into account the error on the metal content of
the calibrating clusters and the photometric calibration errors; (ii)
The relations obtained in this paragraph can be used only in a limited
metal interval and for a limited sample of clusters, which must have a
blue HB.

\section{Summary and discussion \label{sec:sum-discussion}}

In this paper, we took advantage of the homogeneous photometric
databases of Galactic globular clusters, presented in Rosenberg et al
(\cite{dutch00}, \cite{jkt00}) and Piotto et al. (\cite{piottoEtal02}),
to complete another step in the characterization of the morphological
properties of the CMD of GGCs, and in the fine-tuning of the theoretical
models. To this aim, the $W_{\rm HB}$ parameter, originally defined in
B98, has been measured for all suitable clusters, and appropriate
corrections to account for differential reddening effects have been
applied. The quality of the new data sets has allowed to settle the
original question posed in B98. The trend of $W_{\rm HB}$ with
metallicity showed a dispersion that was larger than formal measurement
errors, a fact that can now be attributed to the inhomogeneous data
sources employed.

The dispersion of $W_{\rm HB}$ around the mean trend with metallicity,
is now compatible with the error bar on the data points. This means that
(a) one can reverse the argument and use $W_{\rm HB}$ to have a
first-order guess on a cluster metallicity (Sect.~\ref{sec:metindex}),
and (b) that whatever other variables influence the parameter, they must
have a well-defined dependence on [Fe/H] (including zero dependence).

The second point was investigated in some detail by comparing the data
to two independent sets of theoretical calculations. First, it was
noted that the observed trend of $ W_{\rm HB} $ with metallicity
(which is stronger for the $B-V$ color) is well reproduced by both
isochrone sets, although a zero-point problem in the $(V-I)$
color-temperature transformations seems to be present.

In order to test the dependence on all other variables influencing
$W_{\rm HB}$, their values were varied within reasonable limits. For any
variable $x$, we checked the $W_{\rm HB}$ vs. $x$ trend at several fixed
metallicities, and noticed that these trends can be well approximated
with linear relations. This means that we could use the slope $\Delta
W_{\rm HB}/\Delta x$ to rank the relative dependencies.
We concluded that the influence of the helium content is negligible, and
that that of the mixing-length parameter $\alpha$ is much stronger than
that of the age (see Table~\ref{tab_derivate}). This fact was then
exploited to investigate a long-standing theoretical problem of stellar
evolution, whether $\alpha$ should be varied with metallicity or
not. From our comparisons, {\em we find no trend of $\alpha$ with the
metal content of a cluster}.

With respect to the color-$T_{\rm eff}$ transformations, a test was made
by adopting either the Alonso et al. (\cite{alonsoEtal99}) or the Kurucz
(\cite{kurukz92}) relations. This showed that, {\em potentially}, the
$B-V$ transformations could make a much larger difference than the $V-I$
transformations. However, since actually the theoretical $V-I$ colors
show the greater problems, the agreement between the two
transformations implies that a deep revision for them is needed.

Finally, we examined the question of whether models with enhanced
$\alpha$-elements better reproduce the CMD morphology of GGCs. The
conclusion is that, unfortunately, the existence of two discrepant
metallicity scales leaves this question open. The best results are
obtained in the $\alpha$-enhancement scenario, and adopting the ZW84
metallicity scale. However, one could well adopt the CG97 scale and
claim that the $B-V$ color-transformations are wrong, since we have seen
how much they change from author to author. Thus, we must conclude that
there is an urgent need for an independent metallicity scale (which also
involves the question of galactic chemical evolution models, see Saviane
\& Rosenberg \cite{sr99}).

\begin{acknowledgements}
We acknowledge useful discussions with Eugenio Carretta, that improved
an early draft of this paper. 
This work has been supported by the Ministero della Ricerca
Scientifica e Tecnologica under the program ``Stellar Dynamics and
Stellar Evolution in Globular Clusters'' and by the Agenzia Spaziale
Italiana.
\end{acknowledgements}

\appendix

\section{Differential reddening corrections \label{sec:maths}}

\subsection{Corrections according to Olson \label{sec:olson}}

Olson (\cite{Olso75}) obtained the following relation:\[
R=3.25+0.25\times (B-V)_{0}+0.05\times E_{B-V}\]
>From the definition of $ R=\left( {\frac{A_{V}}{E_{B-V}}}\right)  $ and of
the color excess, we can write: \begin{eqnarray*}
W^{0}_{\rm HB}=(B-V)^{0}_{\rm RGB}-(B-V)_{\rm TD}^{0} & = & \\
=(B-V)_{\rm RGB}-E^{\rm RGB}_{B-V}-(B-V)_{\rm TD}+E^{\rm TD}_{B-V} & = & \\
=W_{\rm obs}+(E^{\rm TD}_{B-V}-E^{\rm RGB}_{B-V}) & = & \\
=W_{\rm obs}+\left( \left. \frac{A_{V}}{R}\right| _{\rm TD}-\left. \frac{A_{V}}{R}\right| _{\rm RGB}\right)  &  & 
\end{eqnarray*}
Since $ A_{V} $ is little dependent on the color, we substitute an average
value in the equation, $ \left< {A_{V}}\right>  $, obtaining:

\[
W_{\rm HB}^{0}=W_{\rm obs}+\left< {A_{V}}\right> \left( {\frac{R^{RGB}-R^{TD}}{R^{TD}\cdot {R^{RGB}}}}\right) .\]
 Now we substitute the product of $ R^{\rm TD} $ times $ R^{\rm RGB} $
by $ R^{2} $, and using Olson expression, we find: \begin{eqnarray*}
W^{0}_{\rm HB}= &  & \\
=W_{\rm obs}+<A_{V}>\left\{ \frac{0.25\, [(B-V)^{\rm RGB}_{0}-(B-V)^{\rm TD}_{0}]}{R^{2}}\right\} = &  & \\
=W_{\rm obs}+\frac{0.25\, E_{B-V}\, \Delta (B-V)_{0}}{R} &  & 
\end{eqnarray*}
Here $ {\textrm{E}_{B-V}}\left( =\frac{\left< A_{V}\right> }{R}\right)  $
is the average color excess of the single cluster, taken from Harris (\cite{harris96}),
and we assumed $ R=3.1 $. As a typical value of $ {\Delta (\textrm{B}-\textrm{V})_{0}} $
we used $ 0.8 $ magnitudes, which is the $ W^{B-V}_{\rm HB} $ value of
NGC~1904, taken from B98 who applied the same differential reddening correction.
In conclusion, the observed values of $ W^{B-V}_{\rm HB} $ must be corrected
with the expression: \begin{equation}
\label{e:o75}
W^{0}_{\rm HB}=W_{\rm obs}+0.064\times E_{B-V}
\end{equation}

\subsection{Corrections according to Grebel and Roberts \label{sec:gr95}}

Grebel and Roberts (\cite{greb95}; GR95) provide tables that give color excesses,
absorptions, and the $ R $ ratio for different photospheric temperatures,
gravities and metallicities ($ {[\textrm{Fe}/\textrm{H}]}=0.0,\, -1.0,\, -2.0 $).
Since we are applying a second-order correction, we used the tables relative
to the intermediate metallicity.

We start again from the relation\begin{equation}
\label{e:whb}
W^{0}_{\rm HB}=W_{\rm obs}=(E^{\rm TD}_{B-V}-E^{\rm RGB}_{B-V})
\end{equation}
And we find the color excesses using the relation\begin{equation}
\label{e:r}
R=\frac{A^{\rm tab}_{V}}{E^{\rm tab}_{B-V}}=\frac{<A_{V}>}{E^{\rm col}_{B-V}}
\end{equation}
where $ A_{V}^{\rm tab} $ and $ {\textrm{E}_{B-V}^{\rm tab}} $ are the
tabular values from GR95, $ {\textrm{E}_{B-V}^{\rm col}} $ is the color excess
for a given cluster and given color (to be determined), and $ \left< {A_{V}}\right>  $
is an average value of the total absorption, which can be calculated with the
expression:\[
<A_{V}>=3.1\times E^{\rm cat}_{B-V}\]
where $ {\textrm{E}_{B-V}^{cat}} $ is the mean reddening taken from Harris
(\cite{harris96}).

Equation (\ref{e:r}) is justified by the fact that the ratio of total to selective
absorption is constant. Thus, it can be used to find $ {\textrm{E}_{B-V}} $
for any TD and RGB color, since: \begin{equation}
\label{e:ebv}
E^{\rm col}_{B-V}=3.1\frac{E^{\rm cat}_{B-V}}{R}
\end{equation}
We now insert Eq.~(\ref{e:ebv}) into Eq.~(\ref{e:whb}), and taking $ R $
from GR95 tables, we obtain\begin{eqnarray*}
W^{0}_{\rm HB}=W_{\rm obs}+3.1\times \left( \frac{1}{R_{\rm TD}}-\frac{1}{R_{\rm RGB}}\right) \times E^{\rm cat}_{B-V} & =
\end{eqnarray*}
\begin{equation}
\label{e:gr95}
=W_{\rm obs}+c\times E^{\rm cat}_{B-V}
\end{equation}
This last equation (\ref{e:gr95}) is formally similar to Eq.~(\ref{e:o75}),
but in this case $ c $ is a variable that depends on the color of the two
CMD points. The variations of $ c $ are, as expected, of the order of a thousandth
of a magnitude.

\subsection{Corrections to the $ {(\textrm{V}-\textrm{I})}$ measurements
\label{sec:vi}}

The analogous of Eq.~(\ref{e:whb}) is the following \begin{equation}
\label{e:whbi}
W^{0}_{\rm HB}=W_{\rm obs}+(E^{\rm TD}_{V-I}-E^{\rm RGB}_{V-I})
\end{equation}
The $ {\textrm{E}_{V-I}} $ values can be computed from $ {\textrm{E}_{B-V}} $
\begin{equation}
\label{e:evi}
E^{\rm col}_{V-I}=\left( \frac{E_{V-I}}{E_{B-V}}\right) ^{\rm tab}\times E^{\rm col}_{B-V}
\end{equation}
where, like before, {}``tab{}'' quantities depend on the temperature and are
taken from GR95 tables, while {}``col{}'' quantities are referred to different
zones of each cluster's CMD. Thus $ {\textrm{E}_{B-V}^{\rm col}} $ is the
value of the blue color excess (Eq.~\ref{e:ebv}), while $ {\textrm{E}_{V-I}^{\rm col}} $
is the value of the color excess that must be computed.

Substituting Eq.~(\ref{e:evi}) into Eq.~(\ref{e:whbi}), and using Eq.~(\ref{e:ebv}),
one obtains: \begin{eqnarray*}
W^{0}_{\rm HB}= &  & \\
=W_{\rm obs}+\left[ \left( \frac{E_{V-I}}{E_{B-V}}\right) ^{\rm tab}_{\rm TD}E^{\rm TD}_{B-V}-\left( \frac{E_{V-I}}{E_{B-V}}\right) ^{\rm tab}_{\rm RGB}E^{\rm RGB}_{B-V}\right] = &  & \\
=W_{\rm obs}+3.1\left[ \left( \frac{1}{R}\frac{E_{V-I}}{E_{B-V}}\right) ^{\rm tab}_{\rm TD}-\left( \frac{1}{R}\frac{E_{V-I}}{E_{B-V}}\right) ^{\rm tab}_{\rm RGB}\right] E^{\rm cat}_{B-V}= &  & \\
=W_{\rm obs}+c'\cdot E^{\rm cat}_{B-V} &  & 
\end{eqnarray*}

\begin{figure}
{\par\centering \resizebox*{0.9\columnwidth}{!}{\includegraphics{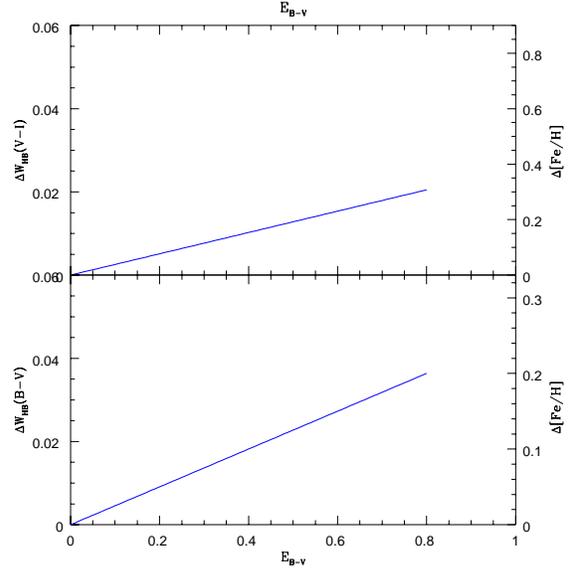}} \par}

\caption{The corrections that must be applied to $ W_{\rm HB}$ are
plotted versus the selective absorption $ E_{B-V}$. Values
relative to $ (B-V)$ color differences are displayed in
the lower panel, while the upper panel displays the corrections to $ (V-I)$
color differences. The right vertical axis shows the impact on {[}Fe/H{]} values
if inferred from $ W_{\rm HB}$ (see Sect.~\ref{sec:metindex}).
\label{fig:and}}
\end{figure}

The corrections to the two $ W_{\rm HB} $ are plotted in Fig.~\ref{fig:and}
as a function of $ E_{B-V} $. It is then clear that $ (V-I) $ colors are
$ \sim 2 $ times less sensitive to this effect. 

\end{document}